\documentclass[a4paper,11pt]{article}
\pdfoutput=1 

\usepackage{jcappub} 

\usepackage[T1]{fontenc} 

\title{Void spin distribution as a powerful probe of $\sigma_{8}$}

\author{Geonwoo Kang}
\author[1]{, Jounghun Lee \note{Corresponding author.}}

\affiliation{Department of Physics and Astronomy, Seoul National University, \\
Kwanak-ro 1, Kwanak-gu, Seoul 08826, Republic of Korea}

\emailAdd{kanggeonwoo@snu.ac.kr}
\emailAdd{cosmos.hun@gmail.com}

\abstract{We present a numerical proof of the concept that the void spin distributions can provide a tight constraint on the amplitude of matter density fluctuation 
on the scale of $8\,h^{-1}{\rm Mpc}$ ($\sigma_{8}$) without being severely deteriorated by the degeneracies of $\sigma_{8}$ with cold dark matter density parameter 
multiplied by the dimensionless Hubble parameter square ($\Omega_{\rm cdm}h^{2}$), total neutrino mass ($M_{\nu}$) and dark energy equation of state ($w$).  
Applying the Void-Finder algorithm~\cite{HV02} to a total of $15$ AbacusSummit $N$-body simulations of $15$ different cosmological models~\cite{summit1}, 
we identify the giant voids and measure the magnitudes of rescaled specific angular momenta of point-like void halos as their spins.  The $15$ cosmologies include the 
Planck $\Lambda$CDM and $14$ non-Planck models, each of which differs among one another only in one of $\{\sigma_{8},\ \Omega_{\rm cdm}h^{2},\ M_{\nu},\ w\}$. 
We determine the probability density distribution of void spins for each model and for the first time find it to be well approximated by the generalized 
Gamma distribution with two characteristic parameters, $k$ and  $\theta$. It turns out that the best-fit values of $k$ and $\theta$ exhibit very sensitive dependence
only on $\sigma_{8}$, being almost insensitive to $\Omega_{\rm cdm}h^{2}$, $M_{\nu}$ and $w$. This exclusive $\sigma_{8}$-dependence of the void spin distributions is 
confirmed to be robust against the variation of the mass and number cuts of void halos. We also test an observational feasibility of estimating the void spins from real data 
on the galaxy redshifts.} 
\begin{document} 
\maketitle
\flushbottom

\section{Introduction}\label{sec:intro}

The persistent and significant tensions between the distant and near field probes on the values of the Hubble constant 
($H_{0}$)~\cite{planck13,rie-etal16,eBoss21,val-etal21}
and standard deviation of the linear density inhomogeneities on the scale of $8\,h^{-1}{\rm Mpc}$ ($\sigma_{8}$)~\cite{jou-etal17,kids450,pal-etal20,asg-etal20,kids1000,desy3,li-etal23} 
have recently drawn considerable attentions~\cite[see][for a recent review]{cosmological_tension}. 
Here, the distant field probe designates the temperature power spectrum of the cosmic microwave background (CMB) radiation, the latest analysis 
of which yielded $H_{0}=67.36\pm 0.54$ and $\sigma_{8}=0.811\pm 0.006$~\cite{planck18}. 
Whereas, the near field counterparts are mainly based on the distance-luminosity relation of  type Ia supernovae (SNIa) and weak lensing shear two-point 
correlation functions, which recently yielded the updated results of $H_{0}=73.04\pm 1.04$~\cite{rie-etal22} and $\sigma_{8}=0.838^{+0.140}_{-0.141}$~\cite{kids21}. 
Although they have yet to blight the legacy of  the {\it concordance} cosmology where the cosmological constant dark energy (DE), $\Lambda$, with 
equation of state $w=-1$ and cold dark matter (CDM) dominate the total energy and matter densities of the universe, respectively. 
These  $H_{0}$ and $\sigma_{8}$ tensions have raised questions about robustness of conventional methodologies and highlighted the 
need for closer scrutiny.

In some literatures, it was suggested that some unknown systematics involved in the near-field probes~\cite{sha-etal19,ben-etal19,mort-etal22,lcdm_well,dha-etal24,efs24} 
should likely be the main cause of the tensions,  given the power of the early-time probes derived from the first principles. In some other literatures, it was argued that given the 
statistical robustness of the late-time probes based on almost model independent non-parameter approaches, 
the $H_{0}$ and $\sigma_{8}$ tensions may hint at {\it new physics}~\cite{cosmological_tension,val-etal21,sch-etal22,kha-etal24}. 
This radical interpretation has recently gained traction by the result of the DE Spectroscopic Instrument (DESI) observations that a dynamical DE model with time-varying 
equation of state, $w(t)$, is preferred to the base $\Lambda$CDM one by the Baryonic Acoustic Oscillation (BAO) data combined with the Planck CMB data~\cite{desi24}. 
Nevertheless, it has been realized that a simple extension of $\Lambda$CDM alone is likely to fail in providing a simultaneous 
solution to the $H_{0}$ and $\sigma_{8}$ tensions~\cite{kee-etal19,jed-etal21,sak23,aka-etal24} and to worsen one of them while solving the other~\cite{cla-etal23,hei-etal23,sun23}, and that 
the $\sigma_{8}$ tension, albeit less significant than the $H_{0}$ tension, should not be neglected in search of a viable new model aimed at improving our understanding of 
cosmic acceleration. In other words, the $\sigma_{8}$ tension ought to be used to constrain further and screen candidate new physics suggested to solve the $H_{0}$ tension~\cite{hei-etal23}.  

In fact, it may require even more meticulous and cautious examinations to resolve the $\sigma_{8}$ tension than the $H_{0}$ counterpart 
given the notorious degeneracy of $\sigma_{8}$ even in the late-time probes with multiple key cosmological parameters such as $w$ 
and total neutrino mass $M_{\nu}\equiv\sum m_{\nu}$ as well as $\Omega_{\rm cdm}h^{2}$ with CDM density parameter $\Omega_{\rm cdm}$ and 
dimensionless Hubble constant $h$~\cite{eke-etal96,WS98,nu_review,hen-etal09}. 
Therefore, before finding a simultaneous solution to both of the tensions, it is quite essential to break the aforementioned cosmological degeneracy 
by developing new diagnostics that exhibit sensitive dependence mainly on $\sigma_{8}$. In fact, much effort has been made over the past two decades 
to develop such non-canonical probes of cosmology and to prove their advantages over the canonical ones likes the two-point 
statistics~\cite[see][for a recent review]{more-etal22,kra-etal24}. 
Here, we put forth the probability density functions of void spins as another non-canonical probe of $\sigma_{8}$, in the hope that it may help us 
break the cosmological degeneracy. 

It has been noted for the past two decades that among the large scale structures, cosmic voids, almost empty large regions surrounded by filaments and sheets, 
exhibit features that make them particularly promising for studying the initial conditions of the universe. Diverse void properties like their abundance as a function of sizes, 
ellipticity distributions, density/velocity profiles, and so forth, have been investigated as possible cosmological 
diagnostics~\cite{LP09,bis-etal10,bos-etal12,LW12,mas-etal15,NT17,ver-etal19,ham-etal20,rez20,dav-etal21,con-etal22,ver-etal23,ebr24}.   
Although these void statistics depend not only on $\sigma_{8}$ but also on $\Omega_{\rm cdm}h^{2}$, $M_{\nu}$ and $w$~\cite{PL07,dav-etal21,con-etal22,ver-etal23,fer-etal24},  
they were found to be useful as complimentary probes of background cosmology since they exhibit orthogonality not only in the $\sigma_{8}$-$\Omega_{\rm cdm}h^{2}$ 
plane~\cite{dav-etal21} but also in constraining $M_{\nu}$ with respect to the other late-time probes like weak gravitational lensing, galaxy clustering and cluster abundance~\cite{ver-etal23}. 

In this paper, we are going to numerically investigate the sensitivity of the probability distributions of void spins, first defined by J.~Lee and D.~Park~\cite{LP06}, to the four 
cosmological parameters, $\sigma_{8}$, $\Omega_{\rm cdm}h^{2}$, $M_{\nu}$ and $w$, and then to assess how feasible it is to estimate this diagnostic in practice. 
The main contents of the upcoming sections are as follows. 
In section~\ref{sec:spin} are presented the descriptions of the numerical dataset, void-identification procedure, and measurements of void spins. 
In section~\ref{sec:model} are presented an analytic formula for the void spin distributions and an explanation for the usage of this formula 
to quantify the dependence of void spin distributions on the aforementioned four parameters.
In section~\ref{sec:sig8} are presented the description of strong $\sigma_{8}$ variation of the void spin distributions. 
In section~\ref{sec:other} is presented the description of the insensitivies of the void spin distributions to $\Omega_{\rm cdm}h^{2}$, $M_{\nu}$ and $w$.  
In section~\ref{sec:obs} is presented a test result of observational feasibility of measuring the void spins from observational data on galaxy redshifts. 
In section~\ref{sec:con} we summarize the results and discuss the limitations of the current analysis and future works to overcome them.

\section{Identification of voids and measurement of their spins}\label{sec:spin}

Our numerical investigation will be based on the halo catalogs from the AbacusSummit~\cite{summit1}, a suite of DM only 
$N$-body simulations conducted for a variety of cosmologies including dynamical DE models with time-varying equation of state ($w$CDM) 
and mixed DM models with massive neutrinos ($\nu\Lambda$CDM). 
The majority of the AbacusSummit simulations was run on a periodic box of a side length $2\,h^{-1}\,{\rm Gpc}$, keeping tracks of $6912^{3}$ DM particles by 
implementing the Abacus code \citep{summit_code}. Compared with the conventional cosmological $N$-body codes, the Abacus code attains unprecedentedly 
high accuracy and rapidity of computing gravities with the help of a new analytical split method for the force decomposition developed by 
M.~V.~L.~Metchnik~\cite{met09}.  
The wide scope of initial conditions for the AbacusSummit simulations were all generated with the help of the {\it Cosmic Linear Anisotropy Solving System} (CLASS)~\cite{class}.

In the AbacusSummit simulations of $\nu\Lambda$CDM models are treated the massive neutrinos as continuous fluid elements, whose suppression effects on the growth of 
structures were incorporated via retracing and rescaling the linear density power spectrum back to $z=99$ from $z=1$~\cite{summit2}. Although no gravitational clustering of massive 
neutrinos with other particle components were properly taken into account, it was claimed to be a good approximation to the neutrino effects on the scales larger than neutrinos free 
streaming lengths at higher redshifts before the onset of their nonlinear evolution~\cite{summit1,summit2}.
Meanwhile, the simulations of $w$CDM models adopted the simplest parametrization of DE equation of state as $w(z)=w_{0}+z\,w_{a}/(1+z)$, assuming no interaction between 
DE and DM. The gravitationally bound DM halos were resolved in each AbacusSummit simulation via the newly developed COMPASO halo-finding 
scheme~\cite{summithalo1,summithalo2}, which is an improved version of the conventional spherical overdensity (SO) halo-finder~\cite{SO}.  
A detailed description of the COMPASO halo-finder and its comparison with the conventional halo-finders are provided in~\cite{summithalo1}. 

For our analysis is chosen a total of $15$ different AbacusSummit simulations, whose initial conditions are different from one another 
only in one of the four cosmological parameters, namely, $\sigma_{8}$, $\Omega_{\rm cdm}h^{2}$, $M_{\nu}$ and $w$. 
The choice of these  $15$ simulations is made through searching for the dataset optimal to the requirements of the current work:
the largest box multiple simulations to explore whether or not the void spin distribution can diminish the notorious degeneracies among these four cosmological 
parameters. Table~\ref{tab:para} lists the values of $\sigma_{8}$, $\Omega_{\rm cdm}h^{2}$, $M_{\nu}$ and $w$, as well as the particle mass resolution ($m_{p}$) 
for the $15$ simulations considered here. The other key cosmological parameters are all set at the same Planck values: $n_{s}=0.9649$ (spectral index) and 
$\Omega_{b}h^{2}=0.024$ (with baryonic matter density parameter $\Omega_{b}$), and $\Omega_{k}=0$ (spatial curvature density parameter). 
The dimensionless Hubble parameter $h$, of each AbacusSummit simulation is set at the value that yields the identical CMB acoustic scale~\cite{summit1}. 

To be consistent with J.Lee and D.Park~\cite{LP06} who for the first time introduced the concept of void spins, we identify voids via the Void-Finder algorithm~\cite{HV02} 
from the halo catalog of each chosen simulation at $z=0$. 
For the usage of this algorithm, it is required to specify two parameters, $s_{c}$ and $l_{c}$, which represent the minimum void-size and wall-field 
halo criterion, respectively~\cite{HV02}. The latter is used to separate the wall halos from the field counterparts, while the former is used to sort out true voids 
from mere gaps among Poisson distribution of halos~\cite{EP97}. The concise summary of the void-identification procedure based on the Void-Finder algorithm 
is provided in the below:
\begin{itemize}
\item
Select the well-resolved DM halos with masses, $m_{h}\ge 10^{11.5}$ in unit of $h^{-1}M_{\odot}$.
For each selected halo,  locate its third nearest neighbor of each selected halo, and measure the separation distance to it, $d_{3}$. 
Take the ensemble average over all selected halos, $\langle d_{3}\rangle $, and compute its standard deviation, 
$\sigma_{3}\equiv [\langle \left(d_{3}-\langle d_{3}\rangle\right)^{2}\rangle]^{1/2}$. 
Determine the value of $l_{c}$ as $\langle d_{3}\rangle + 3\sigma_{3}/2$.   If a halo satisfies the condition of $l_{c}>d_{3}$, then 
classify it as a wall halo.  
\item
Divide the whole simulation box into multiple equal-size grids of side length $l_{c}$.   Locate a block of empty grids, if any, containing no wall halo, and 
determine a spherical volume that fits best the block.  
For each block of empty grids, determine a maximum sphere containing the largest number of empty grids, whose outmost boundary grid just begins to embrace 
three wall halos. 
\item
Rank the volumes of the empty spheres in a decreasing order, and find the largest empty sphere in the box to classify it as a maximal one (a unique void identifier). 
If the volume of a second largest empty sphere does not overlap with that of the largest maximal by more than $10\%$, it is classified as another maximal one. 
Iterate this classification with the lower-ranked empty spheres, as far as their radii are larger than $s_c$.
\item 
Merge the non-maximal spheres into their nearest maximals if their volumes overlap with that of the nearest maximal sphere by more than $50\%$. 
This merged region consisting of the maximal and its overlapping non-maximal spheres is finally identified as a void. 
Using the member halos within the boundary of each void, determine the position and velocity of its center (${\bf x}_{c}$ and ${\bf v}_{c}$, respectively) as 
${\bf x}_{c}\equiv n_{h}^{-1}\sum_{i=1}^{n_{h}}{\bf x}_{i}$ and ${\bf v}_{c}\equiv n_{h}^{-1}\sum_{i=1}^{n_{h}}{\bf v}_{i}$ where ${\bf x}_{i}$ and ${\bf v}_{i}$ are the position and velocity vectors  
of the $i$th void halo, and $n_{h}$ is the number of its member halos. Note that the member void halos are treated as point-like objects with equal masses. 
\item
Compute the effective radius of each void as $R_{\rm eff}\equiv \left(4\pi\,U_{\rm vol}/3\right)^{1/3}$ where the void volume $U_{\rm vol}$ is obtained via the 
Monte-Carlo integration methods~\cite{LP06}. 
The residual over-density of each void, $\delta_{v}$, is also computed as $\delta_{v}\equiv \left(n_{h}-\bar{n}_{h}\right)/\bar{n}_{h}$, where $n_{h}$ and $\bar{n}_{h}$ 
denote the number density of member halos and its mean value, respectively. 
\end{itemize}

\begin{figure}[tbp]
\centering 
\includegraphics[width=0.85\textwidth=0 380 0 200]{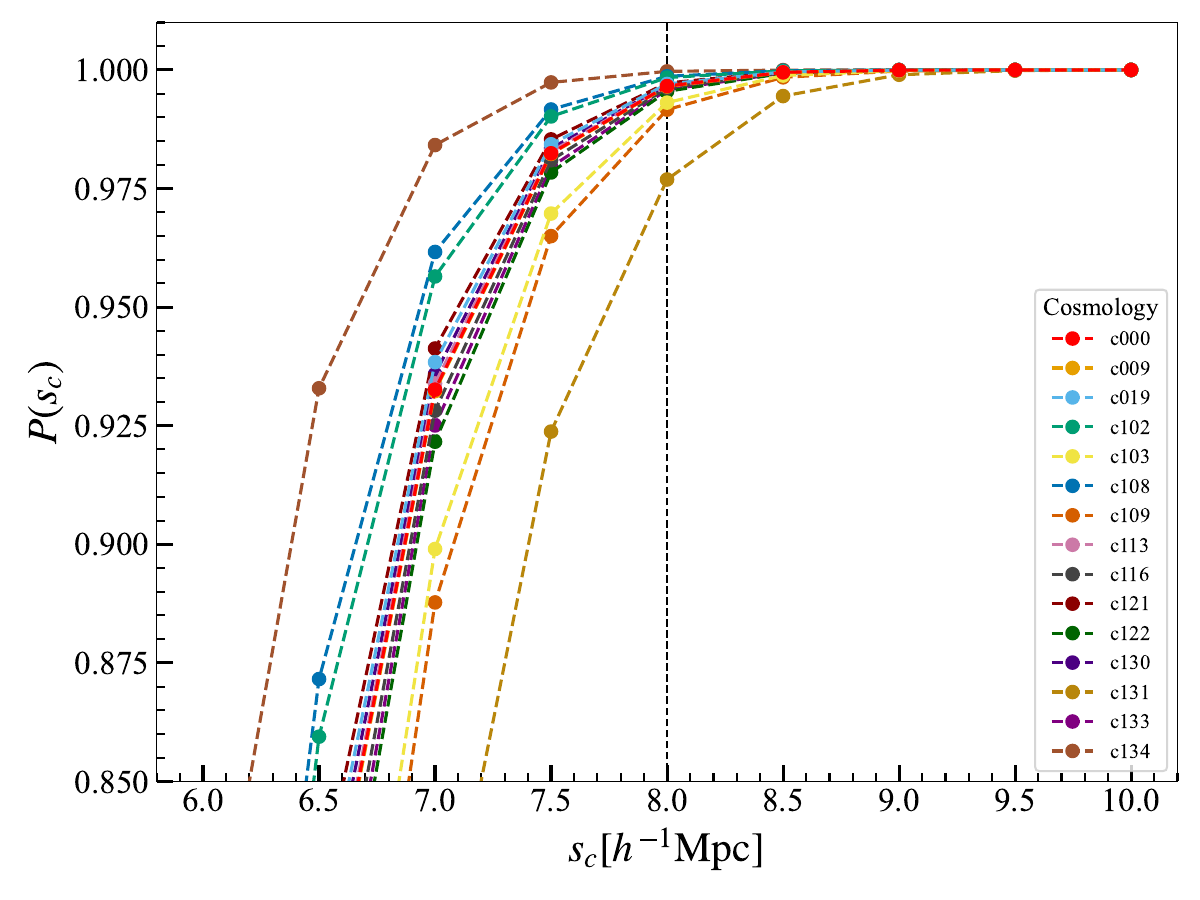}
\caption{\label{fig:sc} Probabilities that the identified empty regions via the Void-Finder algorithm~\cite{HV02} are not mere gaps among halos but true underdense 
voids for $15$ different cosmological models which differ from one another in one of the four cosmological parameters, $\sigma_{8}$, $\Omega_{\rm cdm}h^{2}$, 
$M_{\nu}$ and $w$ (see table~\ref{tab:para}).}
\end{figure}

The total number of voids $N_{v}$, is obviously a function of $s_{c}$.  It is naturally expected that if $s_{c}$ is too small, then mere spatial gaps among the halos 
could be misidentified as voids by the Void-Finder algorithm. To determine a proper value of $s_{c}$ for the identification of genuine voids,  we conduct a statistical significance test 
devised by H.~El-Ad and T.~Piran~\cite{EP97}. Basically, we create $10$ Poisson samples consisting of the same number of halos, and find gaps from 
each sample via the Void-Finder algorithm, and take the average of gap abundance, $N_{\rm gap}(s_{c})$ over the $10$ Poisson samples. 
Finally, the probability defined as $P(s_{c}) = 1 - N_{\rm gap}/N_{\rm v}$, is computed as a function of the minimum void radius $s_{c}$, which is plotted in figure~\ref{fig:sc} for the cases of the $15$ 
AbacusSummit simulations. For all of the $15$ cosmologies considered in the current work, the statistical significances of void abundances exceed $0.95$~\cite{HV02}, 
if $s_{c}\ge 8\,h^{-1}\,{\rm Mpc}$, which leads us to adopt this universal value of $s_{c}=8\,h^{-1}\,{\rm Mpc}$ throughout this work. 

\begin{figure}[tbp]
\centering 
\includegraphics[width=0.85\textwidth=0 380 0 200]{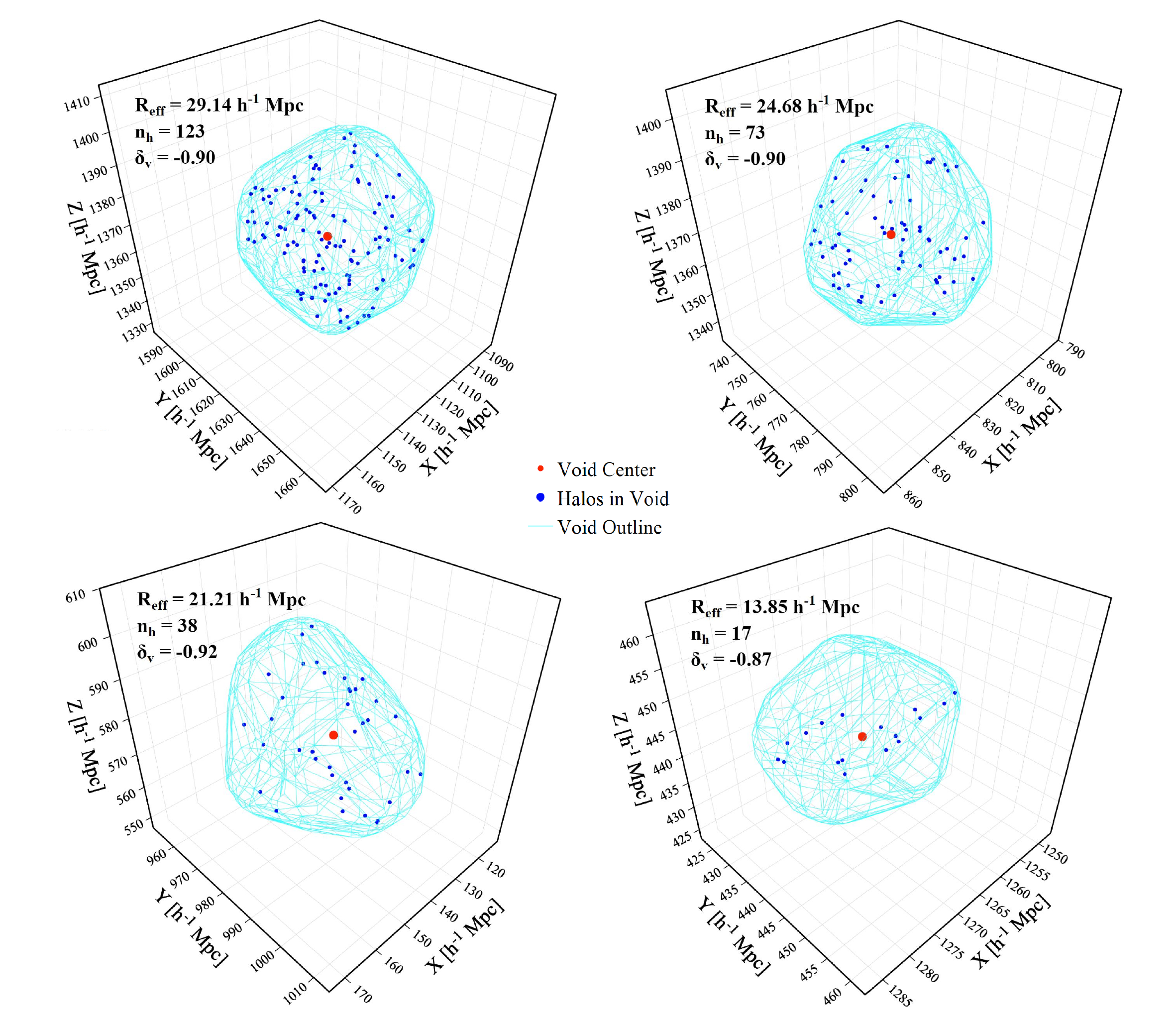}
\caption{\label{fig:void}  Illustration of four example voids identified via the Void-Finder algorithm from the AbacusSummit simulation of the Planck $\Lambda$CDM model 
with effective radii $R_{v}$ and number density contrast of void halos $\delta_{v}$.}
\end{figure}

It is worth mentioning here that the statistical significance test guarantees reliable identifications of genuine voids even in terms of the recently proposed criterion according to which voids 
larger than $2.5$–$3$ times the mean tracer separation are regarded as genuine~\cite{cou-etal19, ham-etal22, con-etal22, ver-etal24}.  For the $\Lambda$CDM case, the mean tracer 
separation turns out to be $1.46\,h^{-1}$Mpc, which indicates that all of the voids in our sample identified with this choice of $s_{c}\ge 8\,h^{-1}\,{\rm Mpc}$ have sizes larger than at least 
$5.5$ times the mean tracer separation. It is found that even for the other cosmological models, this criterion is well satisfied by all of the identified voids.
Examples of four giant voids from the simulation of the Planck $\Lambda$CDM cosmology (c000) are shown in figure~\ref{fig:void}. 
Table~\ref{tab:void} lists the number of voids ($N_{v}$), number of giant voids with $15$ or more member halos, mean effective radii ($\bar{R}_{\rm eff}$), 
mean halo number density ($\bar{\delta}_{h}$) and the wall-to-field halo criterion values ($l_{c}$). 

It should be also worth explaining here why we stick with the classical Void-Finder algorithm rather than the more recently developed ones such as the spherical under-density 
algorithms (e.g., Pylians, Sparkling, Popcorn)~\cite{ban-etal16, pad-etal05,rui-etal15,paz-etal23} and watershed void identification methods (e.g., Zobov, Vide and Revolver)~\cite{zobov,vide,rev}.
True as it is that those advanced algorithms have various advantages and merits over the Void-Finder,  they may not necessarily be the most optimal ones for the determination of 
void spin distributions. In particular, the primary motivation for our choice of the Void-Finder is the result of the recent observational analysis~\cite{LM23} that the signals of the large-scale 
tidal effects on the void galaxies were detected at higher confidence levels when the voids had been identified via the Void-Finder rather than via the watershed counterparts 
(Revolver and Vide). By the same token, we are concerned about the fact that it has yet to be examined whether or not the {\it spherical} under-density void-finding algorithms 
would be more effective than the Void-Finder in capturing the anisotropic tidal effects on the void galaxies, which are essential for the generation of the spinning motions of voids. 
These concerns lead us to make a safety choice, the Void-Finder, whose effectiveness for the determination of void spin distribution was already verified by J.Lee and D.Park~\cite{LP06}. 

\begin{table}[tbp]
\centering
\begin{tabular}{cccccc}
\hline
\hline
\rule{0pt}{4ex}\noindent
model & $\sigma_{8}$ & $\Omega_{\rm cdm} h^2$ & $M_{\nu}$ & $w$ & $m_{p}$ \medskip\\
      &              &                        & $[{\rm eV}]$ &            & [$10^{9}\,h^{-1}\,M_\odot$] \medskip\\
\hline
\rule{0pt}{4ex}\noindent
$c000$ & $0.811$ & $0.120$ & $0.06$  & $-1.0$   & $2.00$ \medskip\\
\hline
\rule{0pt}{4ex}\noindent
$c113$ & $0.795$ & $0.120$ & $0.06$  & $-1.0$   & $2.00$ \medskip\\
$c116$ & $0.869$ & $0.120$ & $0.06$  & $-1.0$   & $2.00$ \medskip\\
$c130$ & $0.714$ & $0.120$ & $0.06$  & $-1.0$   & $2.00$ \medskip\\
$c133$ & $0.908$ & $0.120$ & $0.06$  & $-1.0$   & $2.00$ \medskip\\
\hline
\rule{0pt}{4ex}\noindent
$c102$ & $0.811$ & $0.124$ & $0.06$  & $-1.0$   & $2.06$ \medskip\\
$c103$ & $0.811$ & $0.116$ & $0.06$  & $-1.0$   & $1.94$ \medskip\\
$c131$ & $0.811$ & $0.108$ & $0.06$  & $-1.0$   & $1.83$ \medskip\\
$c134$ & $0.811$ & $0.132$ & $0.06$  & $-1.0$   & $2.17$ \medskip\\
\hline
\rule{0pt}{4ex}\noindent
$c009$ & $0.811$ & $0.120$ & $0.00$  & $-1.0$   & $2.00$ \medskip\\
$c019$ & $0.811$ & $0.120$ & $0.12$  & $-1.0$   & $2.00$ \medskip\\
\hline
\rule{0pt}{4ex}\noindent
$c108$ & $0.811$ & $0.120$ & $0.06$  & $-0.9$   & $2.00$ \medskip\\
$c109$ & $0.811$ & $0.120$ & $0.06$  & $-1.1$   & $2.00$ \medskip\\
$c121$ & $0.811$ & $0.120$ & $0.06$  & $-0.975$ & $2.00$ \medskip\\
$c122$ & $0.811$ & $0.120$ & $0.06$  & $-1.025$ & $2.00$ \medskip\\
\hline
\end{tabular}
\caption{\label{tab:para}
Model, amplitude of the CDM density parameter multiplied by dimensionless Hubble constant square, total neutrinos mass, DE equation of state, 
and mass resolution.}
\end{table}
\begin{table}[tbp]
\centering
\begin{tabular}{cccccc}
\hline
\hline
\rule{0pt}{4ex}\noindent
model & $N_{v}$ & $N_{v}(n_{h} \geq 15)$ & $\bar{R}_{\rm eff}$ & $\bar{\delta_v}$ & $l_c$ \medskip\\
& & & [$h^{-1}$Mpc]  & & [$h^{-1}$Mpc] \medskip\\
\hline
\rule{0pt}{4ex}\noindent
c000 & 633932 & 188804 & 15.20 & -0.87 & 4.50 \medskip\\
\hline
\rule{0pt}{4ex}\noindent
c113 & 633800 & 188561 & 15.18 & -0.87 & 4.50 \medskip\\
c116 & 636195 & 187907 & 15.28 & -0.87 & 4.52 \medskip\\
c130 & 633349 & 188399 & 15.12 & -0.87 & 4.49 \medskip\\
c133 & 637702 & 186933 & 15.34 & -0.87 & 4.54 \medskip\\
\hline
\rule{0pt}{4ex}\noindent
c102 & 616167 & 201465 & 14.80 & -0.87 & 4.40 \medskip\\
c103 & 649396 & 176263 & 15.62 & -0.87 & 4.62 \medskip\\
c131 & 671979 & 153260 & 16.51 & -0.87 & 4.86 \medskip\\
c134 & 572507 & 228304 & 14.07 & -0.87 & 4.19 \medskip\\
\hline
\rule{0pt}{4ex}\noindent
c009 & 634499 & 188468 & 15.20 & -0.87 & 4.51 \medskip\\
c019 & 630400 & 191454 & 15.11 & -0.87 & 4.48 \medskip\\
\hline
\rule{0pt}{4ex}\noindent
c108 & 612243 & 205815 & 14.73 & -0.87 & 4.37 \medskip\\
c109 & 653567 & 171388 & 15.72 & -0.87 & 4.66 \medskip\\
c121 & 628681 & 193031 & 15.07 & -0.87 & 4.47 \medskip\\
c122 & 640020 & 183450 & 15.34 & -0.87 & 4.55 \medskip\\
\hline
\end{tabular}
\caption{\label{tab:void}
Total number of voids, number of giant voids having $\geq$ 15 halos, mean effective radii,
mean density contrast and criterion distance for the classification of wall halos.}
\end{table}

The dimensionless spin vector, ${\bf j}$,  of an identified void is determined as its specific angular momentum: 
\begin{equation}
\label{eqn:def_vspin}
{\bf j}\equiv \frac{1}{\sqrt{2}(V_{v}R_{\rm eff})}\sum_{i=1}^{n_{h}}\left({\bf x}_{i}-{\bf x}_{c}\right)\times \left({\bf v}_{i}-{\bf v}_{c}\right)\, .
\end{equation}
where $V_{v}$ is the mean circular velocity at $R_{\rm eff}$. Note that unlike in the original definition of void spins~\cite{LP06}, a contribution of the $i$th 
member  halo to ${\bf j}$ is not weighted by it mass, which is consistent with the way in which we determine ${\bf x}_{c}$ and ${\bf v}_{c}$.
Given that the virial masses of individual void halos are quite difficult to measure with high accuracy 
in practice, we use this modified definition of void spin vectors and centers. 
Hereafter, the magnitude of this void spin vector, $j\equiv \vert{\bf j}\vert$, will be referred to as a void spin, as in~\cite{LP06}. 
Eq.~(\ref{eqn:def_vspin}) indicates that it is not possible to measure a void spin if $n_{h}$ is too low. The larger number of halos a void has, 
the more accurately its spin can be measured. In accordance, we apply a halo number cut, $n_{c}=15$, to the identified voids from each simulation 
and exclude those voids with $n_{h}< n_{c}$ for the determination of void spin distribution.

\section{Dependence of the void spin distribution on the initial condition}

\subsection{An analytic formula for the void spin distribution}\label{sec:model}
\begin{figure}[tbp]
\centering 
\includegraphics[width=0.85\textwidth=0 380 0 200]{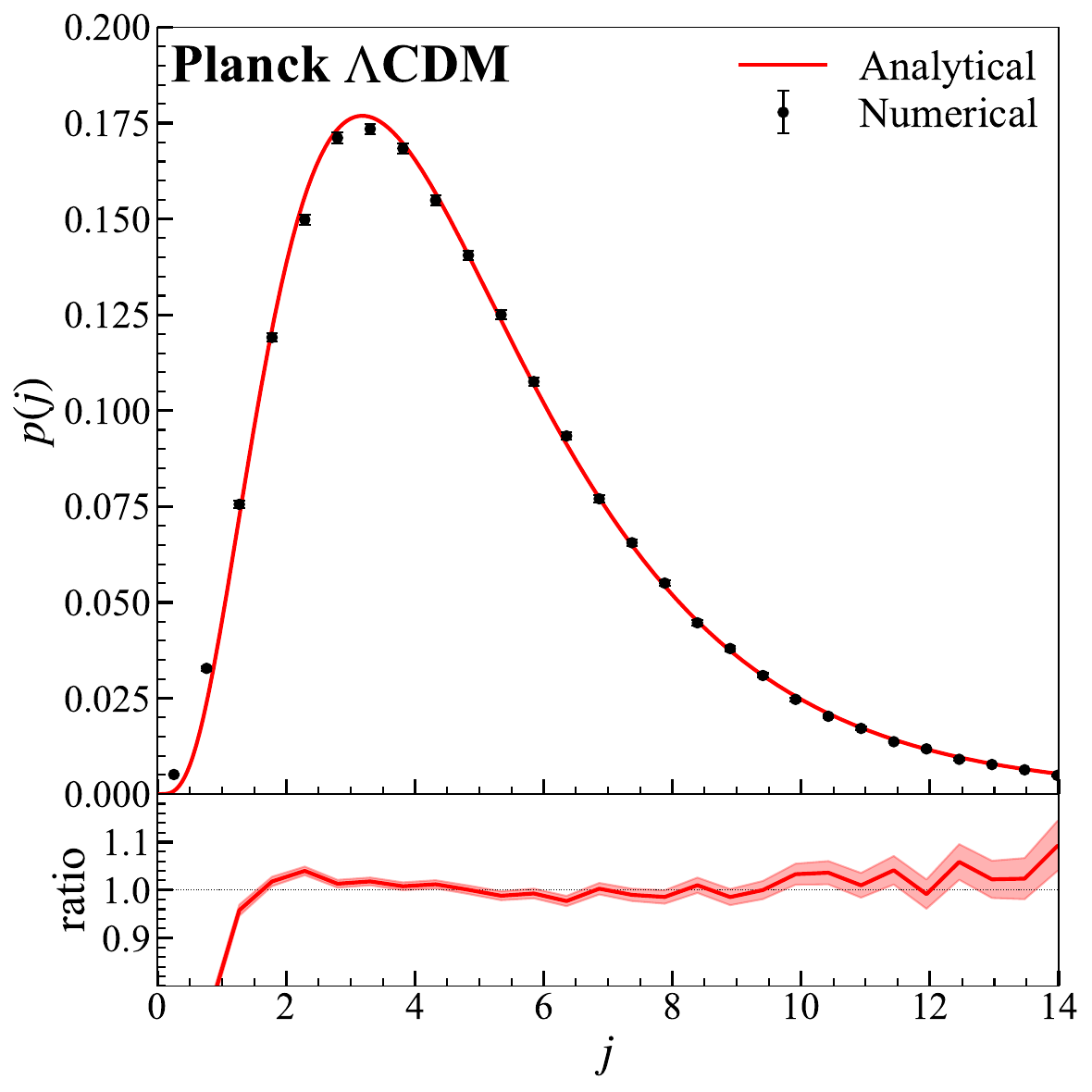}
\caption{\label{fig:pj}  (Top panel): probability density distribution of void spins (black filled circles) with Poisson errors 
from the AbacusSummit simulations of the Planck $\Lambda$CDM cosmology (c000) and the generalized Gamma distribution with the 
best-fit parameters (red solid lines); (Bottom panel): ratio of the numerically obtained $p(j)$ to its best-fit analytical model (red solid lines) 
with $1\sigma$ uncertainties (shaded regions).}
\end{figure}
\begin{figure}[tbp]
\centering 
\includegraphics[width=0.85\textwidth=0 380 0 200]{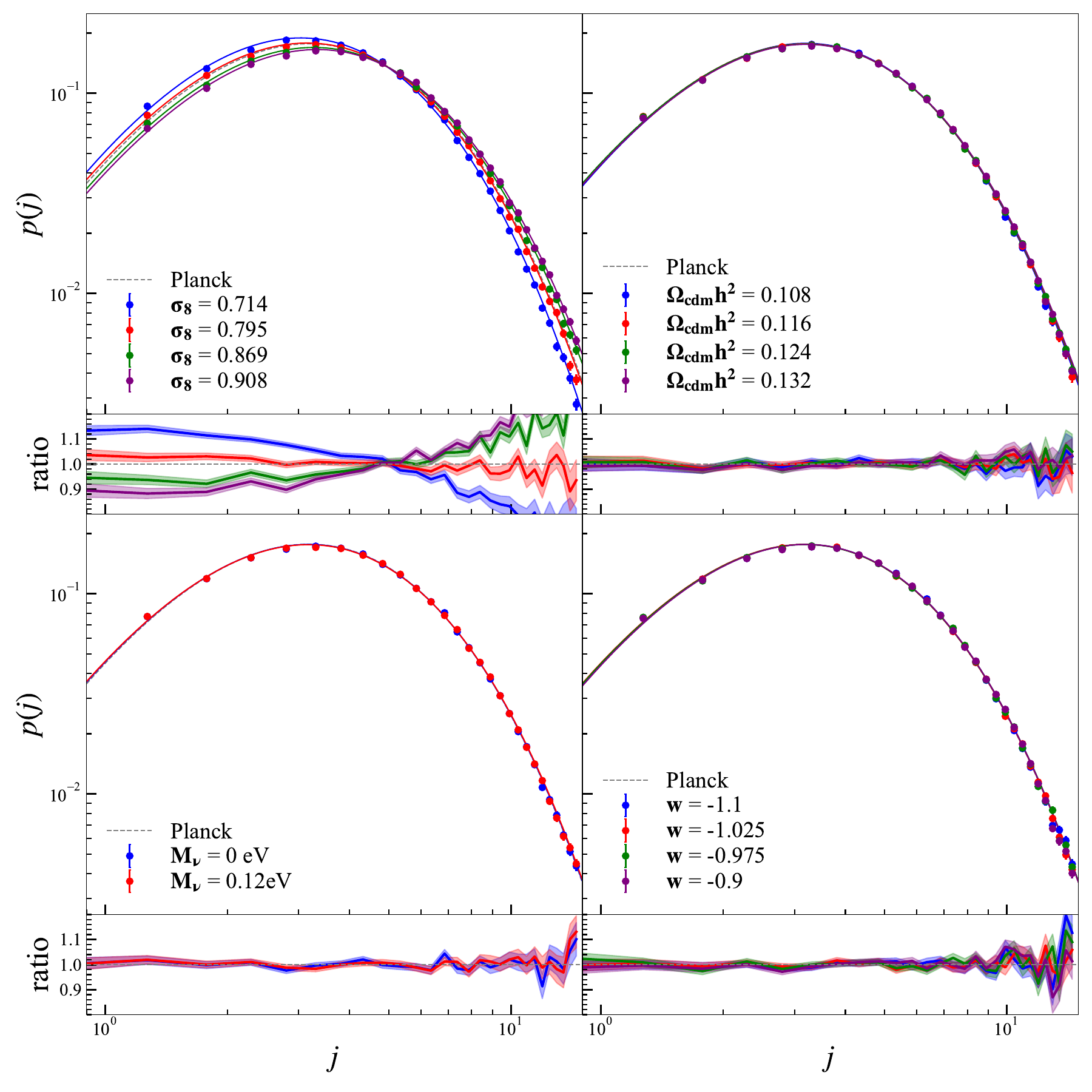}
\caption{\label{fig:loglog}  Comparison of the void spin distributions among $4$ different cases of $\sigma_{8}$ (top-left panel), 
among four different cases of $\Omega_{\rm cdm}h^{2}$ (top-right panel), among two different cases of $M_{\nu}$ (bottom-right panel)
and among four different cases of $w$ (bottom-left panel). In the lower sub-panel of each panel are shown the ratios of the void spin distributions
of the four non-Planck cosmologies to that of the Planck $\Lambda$CDM case.} 
\end{figure}

Splitting the range of $j$ into multiple short intervals of equal length, $\Delta\,j$, and counting the number, $\Delta N_{v}$, of voids whose values of $j$ 
fall in each interval, we determine the probability density distribution of void spins as $p(j) \equiv \Delta\,N_{v}/(N_{v}\Delta\,j)$. 
The top panel of figure~\ref{fig:pj} plots $p(j)$ as black filled circles with Poisson errors for the case of the Planck cosmology (c000). As can be seen, 
just like the well-known spin parameter distribution of galactic halos~\cite{bul-etal01}, the void spin distribution, $p(j)$, shows a long high-$j$ tail. 
Recalling the recent work of J.~Moon and J.~Lee~\cite{ML24} where the spin parameter distribution of DM halos was found to be well approximated by the 
Gamma distribution, we compare the numerically obtained $p(j)$ to the following {\it generalized} Gamma distribution: 
\begin{equation}
\label{eqn:pj}
p(j) = \frac{j^{k-1}}{2\Gamma\left(2k\right)\theta^{k}}\exp\left[-\left(\frac{j}{\theta}\right)^{1/2}\right] \, ,
\end{equation}
where $\Gamma(2k)\equiv \int_{0}^{\infty}\,dt\,t^{2k}e^{-t}$, and $\{k,\ \theta\}$ are two adjustable parameters.
The best-fit values of $\{k,\ \theta\}$ in eq.~(\ref{eqn:pj}) are determined by adjusting the formula to the the numerically obtained $p(j)$ with the help of the $\chi^{2}$-statistics. 
The top and bottom panels of figure~\ref{fig:pj} plot the best-fit analytic formula and its ratio to the numerically obtained $p(j)$ with $1\sigma$ errors 
(red solid lines with shaded area), respectively, for the Planck $\Lambda$CDM model. 
As can be seen, the generalized Gamma distribution gives a good match to the numerical results in almost entire range of $j$ except for the lowest-$j$ bin ($j<1$), 
verifying the validity of eq.~(\ref{eqn:pj}). The relatively large disagreements between the analytical and numerical $p(j)$ in the range of $j< 1$ must reflect the fact that 
the member halos of those voids with low spins ($j<1$) must have almost isotropic spatial distributions, which are likely to cause large uncertainties in the measurements of $j$.  

\subsection{Dependence of the void spin distribution on $\sigma_{8}$}\label{sec:sig8}

\begin{figure}[tbp]
\centering 
\includegraphics[width=0.85\textwidth=0 380 0 200]{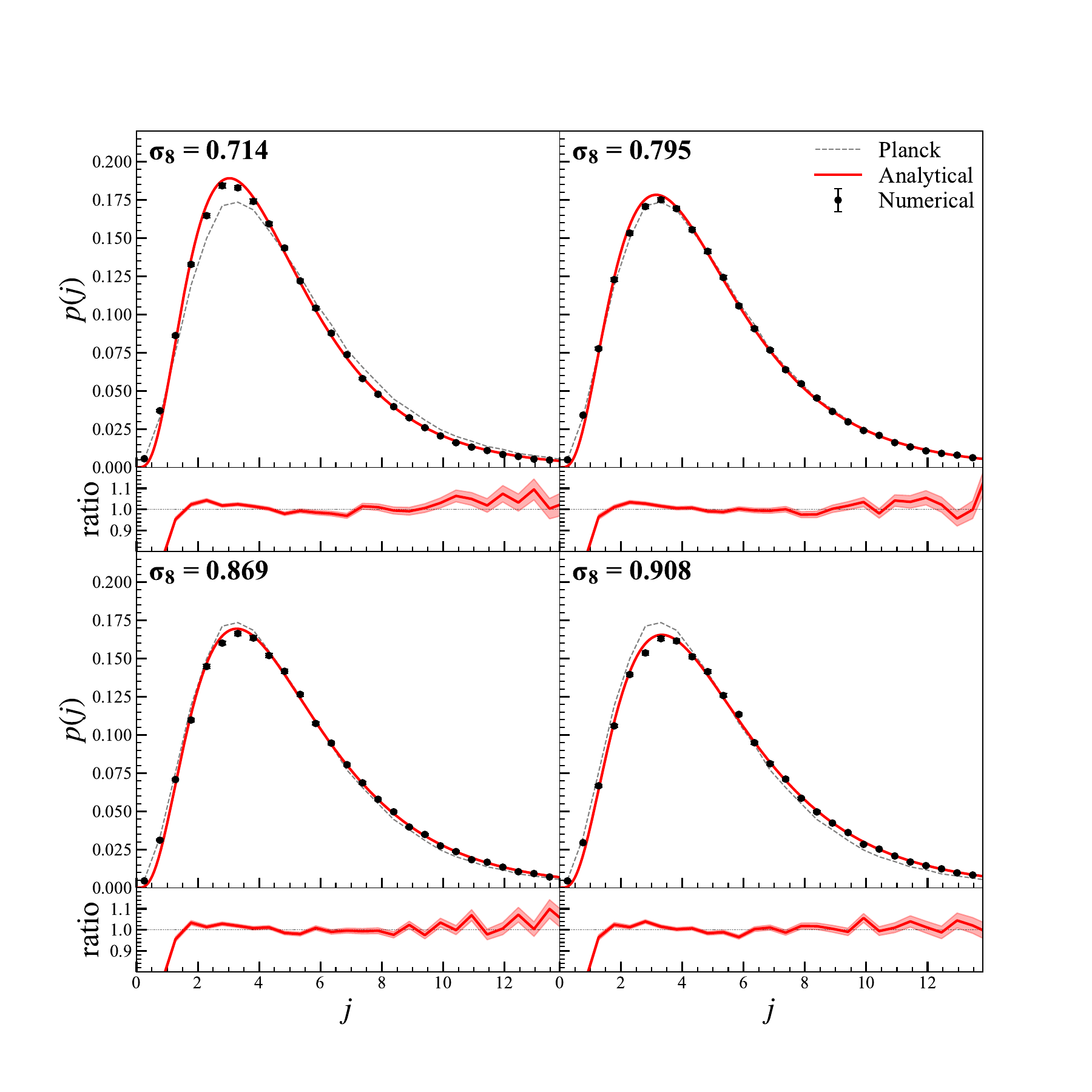}
\caption{\label{fig:pj_sig8} Same as figure~\ref{fig:pj} but for four different cases of $\sigma_{8}$. The void spin distribution for the Planck $\Lambda$CDM 
case is shown as dashed line in each panel for comparison.}
\end{figure}

\begin{figure}[tbp]
\centering 
\includegraphics[width=0.85\textwidth=0 380 0 200]{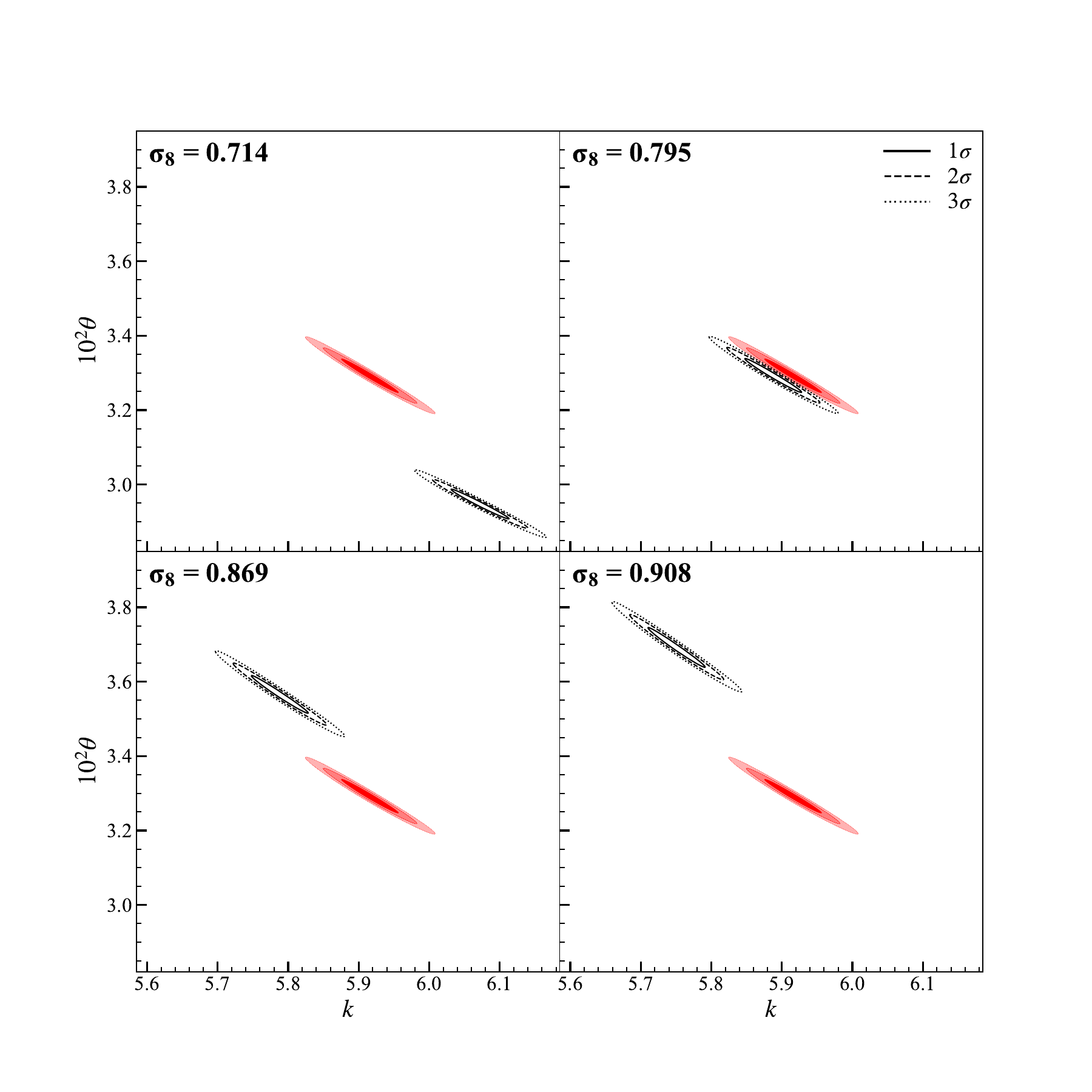}
\caption{\label{fig:cont_sig8} Contours of $68\%$, $95\%$ and $99\%$ confidence area in the two dimensional configuration space spanned 
by $k$ and $\theta$ for the five different cases of $\sigma_{8}$. In each panel, the black unfilled contours correspond to the non-Planck $\Lambda$CDM case 
with a given $\sigma_{8}$ while the red filled contours correspond to the Planck case.}
\end{figure}
\begin{figure}[tbp]
\centering 
\includegraphics[width=0.85\textwidth=0 380 0 200]{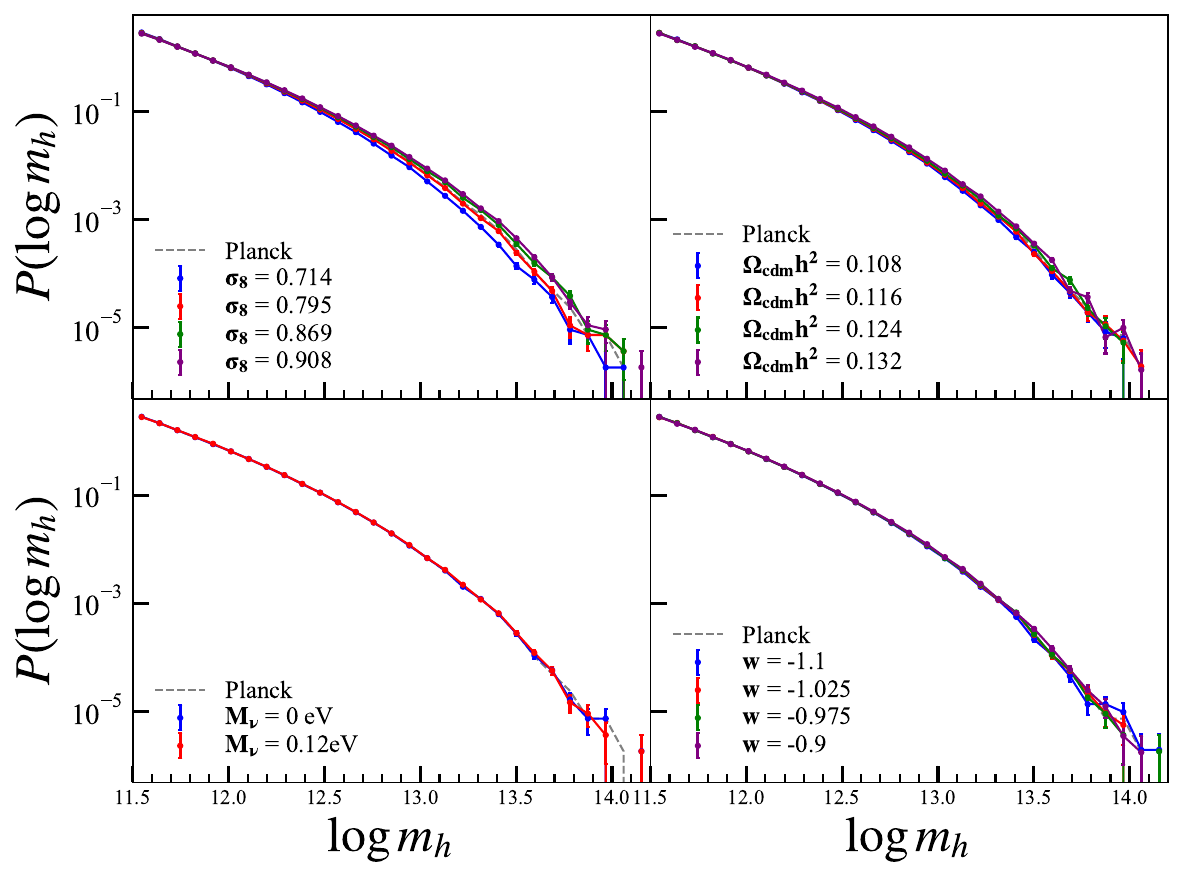}
\caption{\label{fig:mass_dis} Probability density distributions of void halo masses in the logarithmic scale for four different cases of $\sigma_{8}$ 
(top-left panel), for four different cases of $\Omega_{\rm cdm}h^{2}$ (top-right panel), for two different cases of $M_{\nu}$ (bottom-right panel)
and for four different cases of $w$ (bottom-left panel).}
\end{figure}
\begin{figure}[tbp]
\centering 
\includegraphics[width=0.85\textwidth=0 380 0 200]{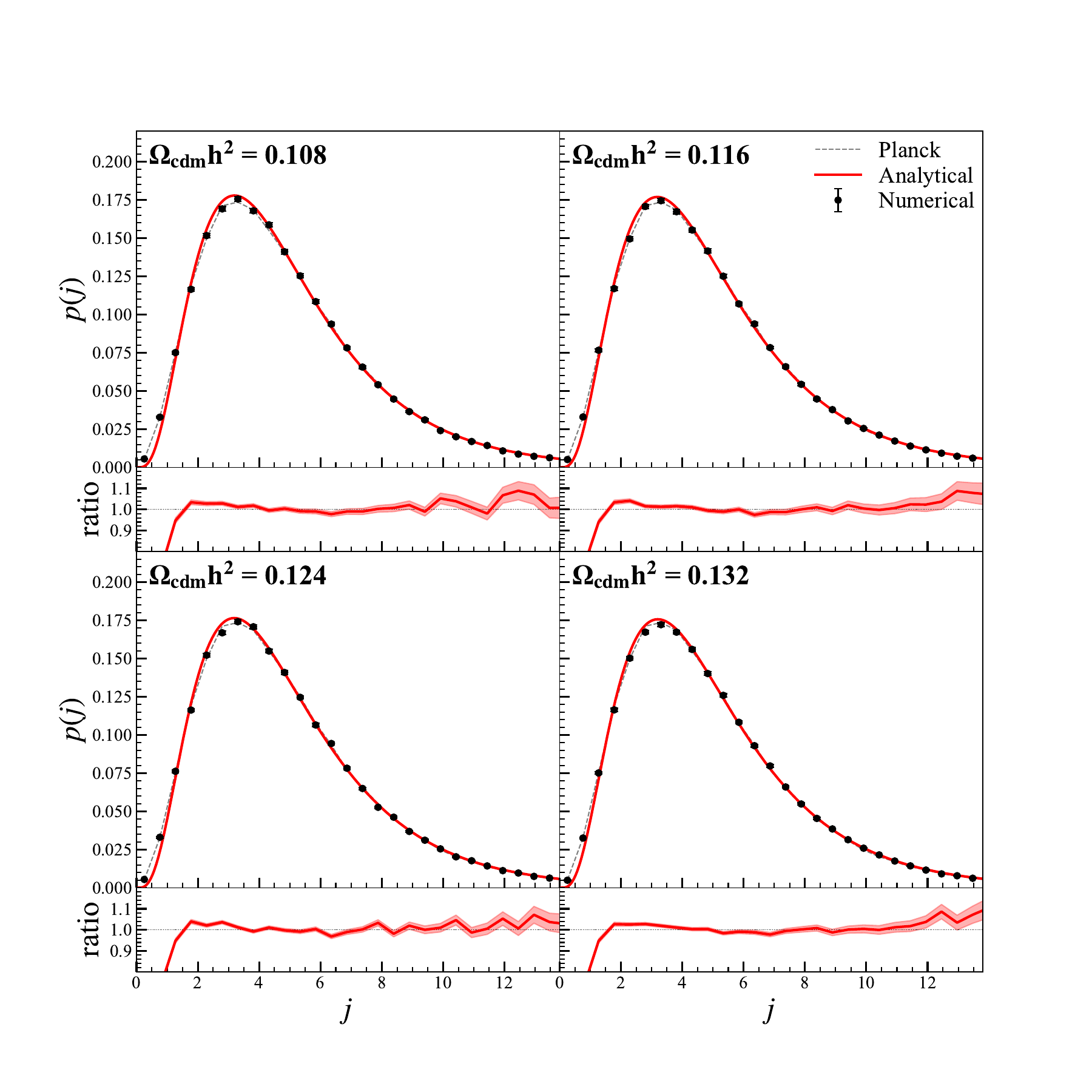}
\caption{\label{fig:pj_om} Same as figure~\ref{fig:pj_sig8} but for four different cases of $\Omega_{\rm cdm}h^{2}$. The void spin distribution for the 
Planck $\Lambda$CDM case is shown as dashed line in each panel for comparison.}
\end{figure}
\begin{figure}[tbp]
\centering 
\includegraphics[width=0.85\textwidth=0 380 0 200]{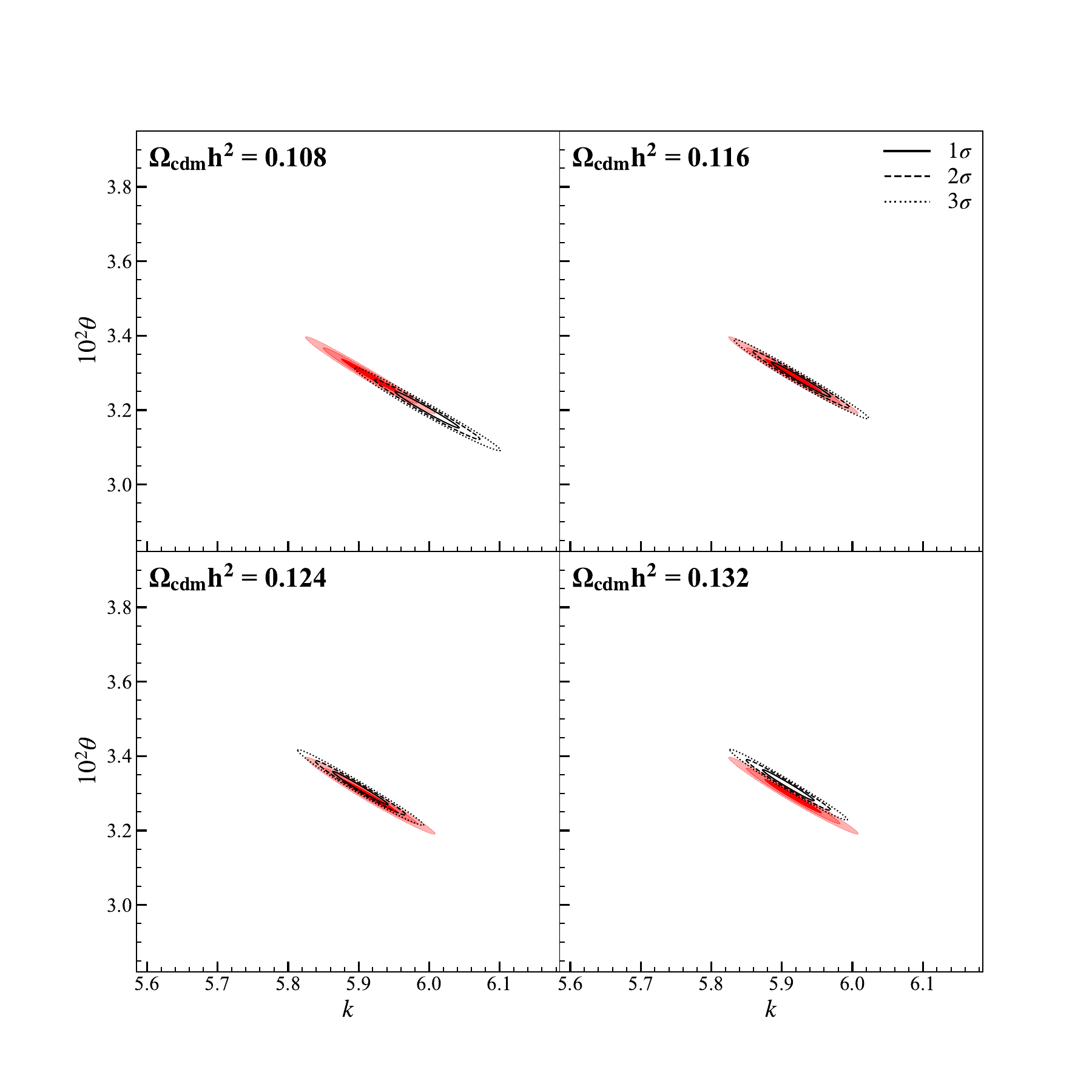}
\caption{\label{fig:cont_om} Same as figure~\ref{fig:cont_sig8} but for four different cases of $\Omega_{\rm cdm}h^{2}$.}
\end{figure}
\begin{figure}[tbp]
\centering 
\includegraphics[width=0.85\textwidth=0 380 0 200]{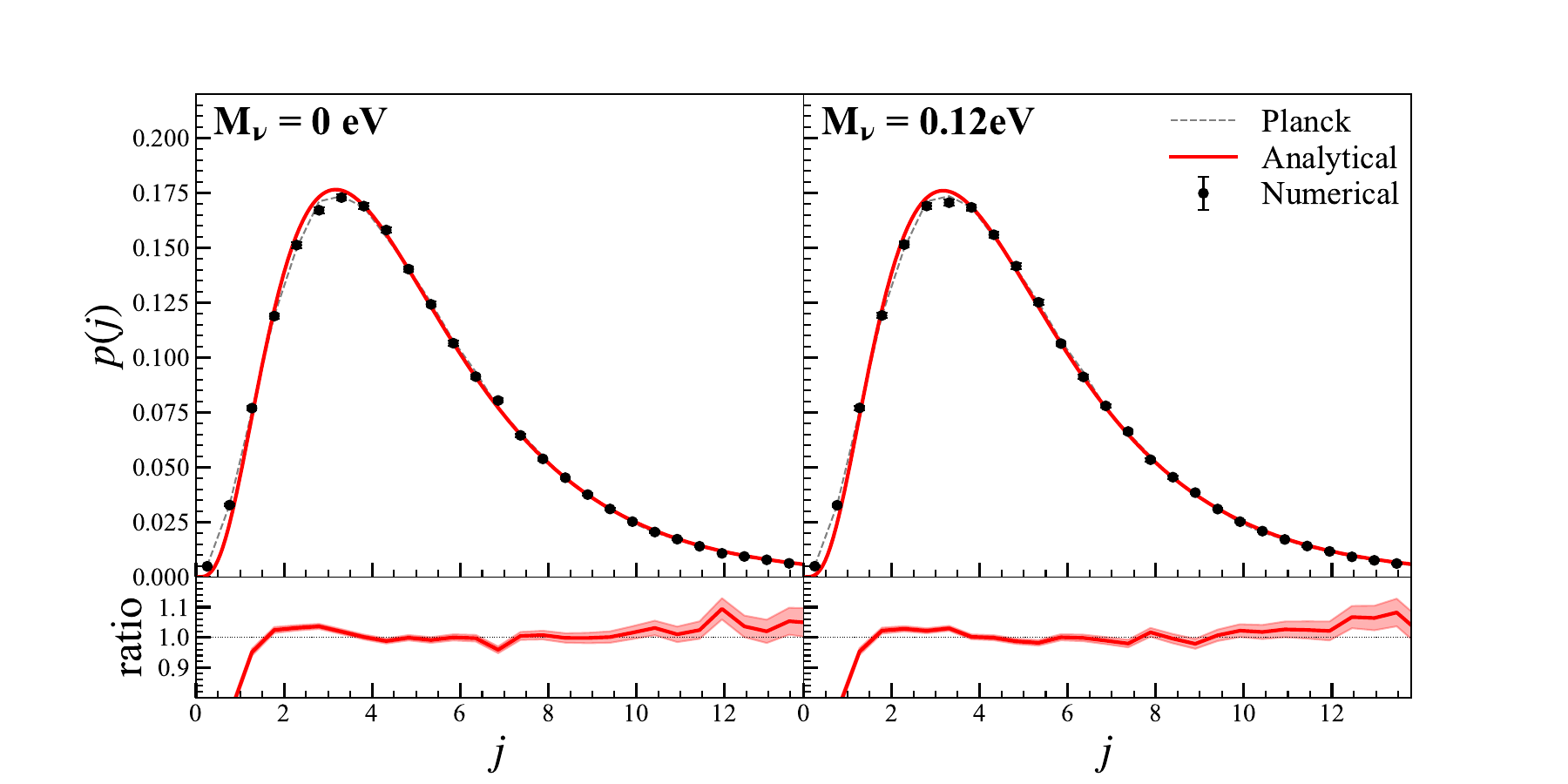}
\caption{\label{fig:pj_nu} Same as figure~\ref{fig:pj_sig8} but for two different cases of $M_{\nu}$. 
The void spin distribution for the Planck $\Lambda$CDM case is shown as dashed line in each panel for comparison.}
\end{figure}
\begin{figure}[tbp]
\centering 
\includegraphics[width=0.85\textwidth=0 380 0 200]{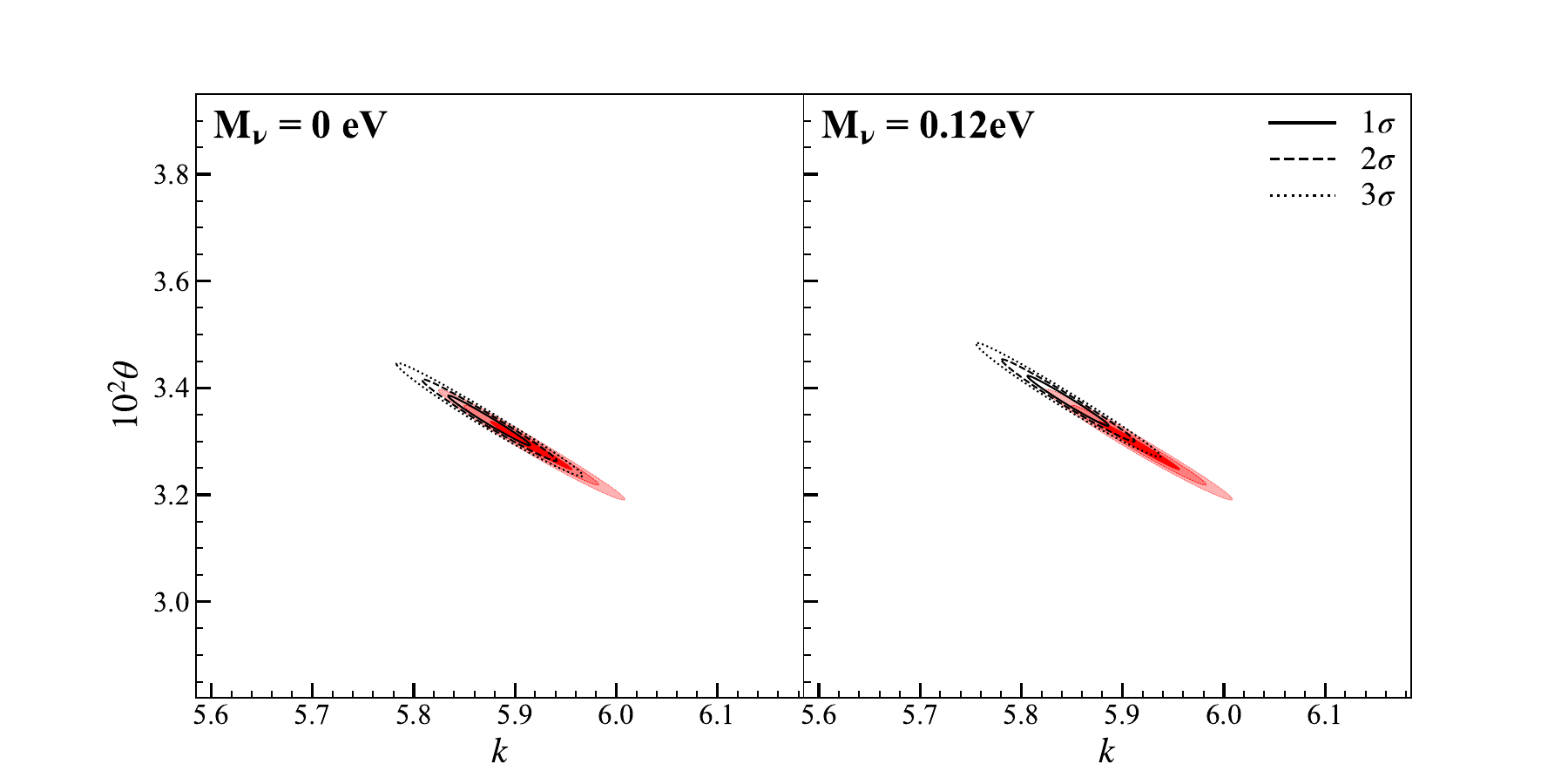}
\caption{\label{fig:cont_nu} Same as figure~\ref{fig:cont_sig8} but for two different cases of $M_{\nu}$.}
\end{figure}
\begin{figure}[tbp]
\centering 
\includegraphics[width=0.85\textwidth=0 380 0 200]{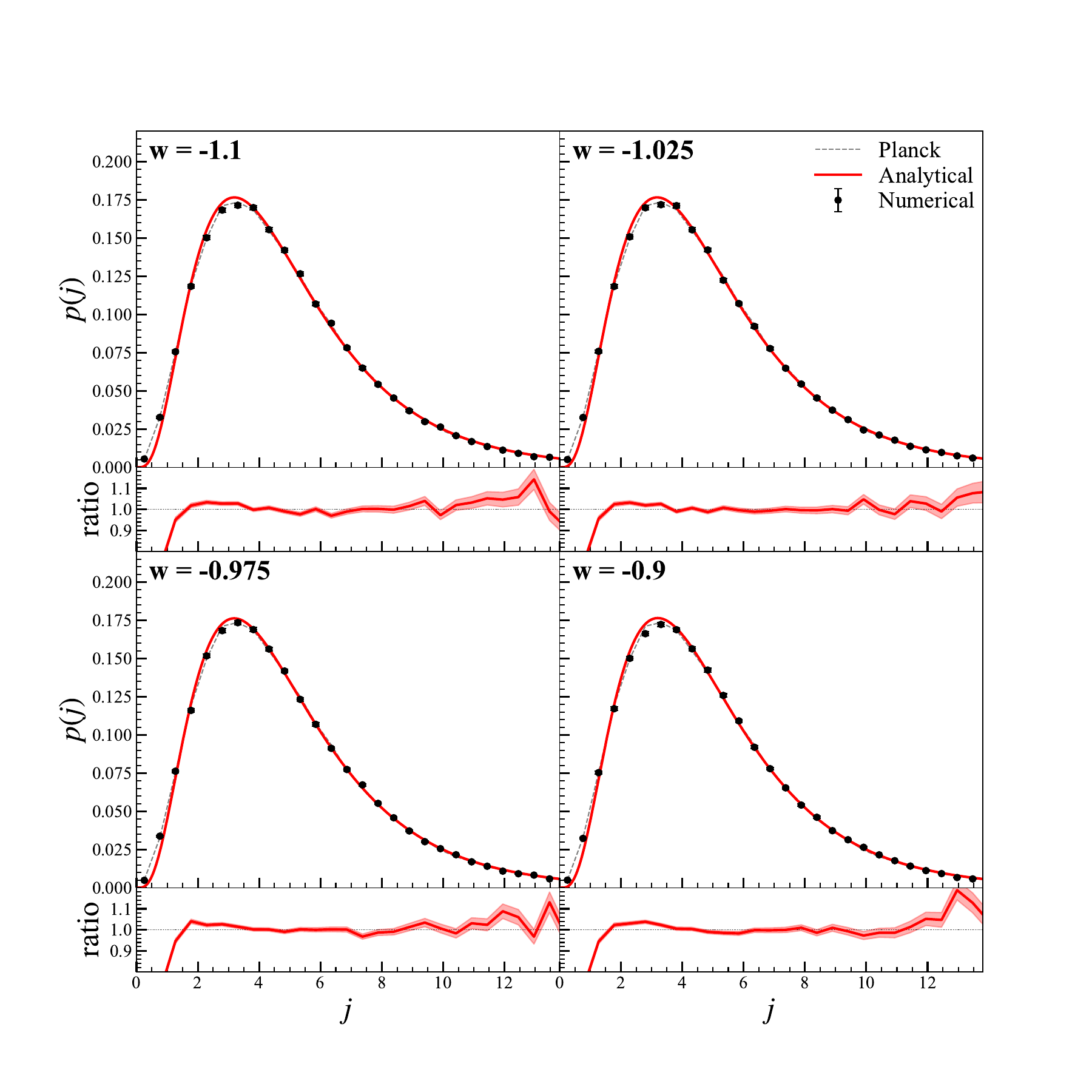}
\caption{\label{fig:pj_w} Same as figure~\ref{fig:pj_sig8} but for four different cases of $w$. 
The void spin distribution for the Planck $\Lambda$CDM case is shown as dashed line in each panel for comparison.}
\end{figure}
\begin{figure}[tbp]
\centering 
\includegraphics[width=0.85\textwidth=0 380 0 200]{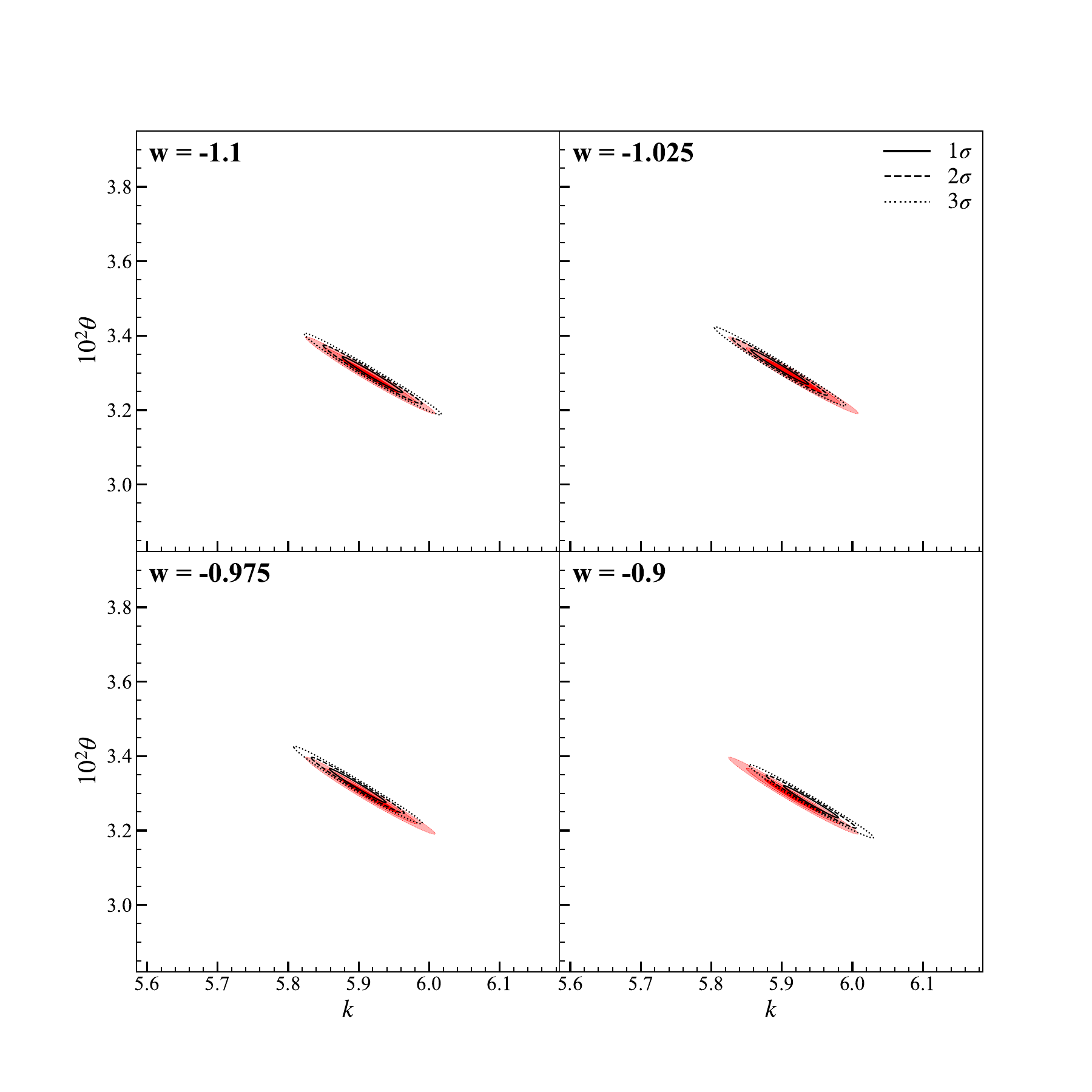}
\caption{\label{fig:cont_w} Same as figure~\ref{fig:cont_sig8} for four different cases of $w$.}
\end{figure}
\begin{table}[tbp]
\centering
\begin{tabular}{cccc}
\hline
\hline
\rule{0pt}{4ex}\noindent
model & $k$ & $10^{2}\theta$ & $D/\sigma_{D}$ \medskip\\
\hline
\rule{0pt}{4ex}\noindent
c000 & $5.92^{+0.026}_{-0.027}$ & $3.29^{+0.0305}_{-0.0294}$ & $0.00$ \medskip\\
\hline
\rule{0pt}{4ex}\noindent
c113 & $5.89^{+0.026}_{-0.026}$ & $3.29^{+0.0305}_{-0.0294}$ & $0.75$ \medskip\\
c116 & $5.79^{+0.026}_{-0.026}$ & $3.57^{+0.0339}_{-0.0328}$ & $3.39$ \medskip\\
c130 & $6.07^{+0.028}_{-0.026}$ & $2.95^{+0.0260}_{-0.0271}$ & $4.08$ \medskip\\
c133 & $5.75^{+0.027}_{-0.026}$ & $3.69^{+0.0351}_{-0.0351}$ & $4.38$ \medskip\\
\hline
\rule{0pt}{4ex}\noindent
c102 & $5.90^{+0.025}_{-0.026}$ & $3.31^{+0.0305}_{-0.0283}$ & $0.35$ \medskip\\
c103 & $5.93^{+0.027}_{-0.028}$ & $3.28^{+0.0317}_{-0.0305}$ & $0.30$ \medskip\\
c131 & $6.00^{+0.030}_{-0.029}$ & $3.20^{+0.0317}_{-0.0328}$ & $2.00$ \medskip\\
c134 & $5.91^{+0.025}_{-0.024}$ & $3.32^{+0.0271}_{-0.0283}$ & $0.20$ \medskip\\
\hline
\rule{0pt}{4ex}\noindent
c009 & $5.87^{+0.027}_{-0.026}$ & $3.34^{+0.0305}_{-0.0305}$ & $1.11$ \medskip\\
c019 & $5.85^{+0.027}_{-0.026}$ & $3.38^{+0.0305}_{-0.0317}$ & $1.88$ \medskip\\
\hline
\rule{0pt}{4ex}\noindent
c108 & $5.94^{+0.026}_{-0.025}$ & $3.28^{+0.0283}_{-0.0283}$ & $0.68$ \medskip\\
c109 & $5.92^{+0.027}_{-0.029}$ & $3.29^{+0.0328}_{-0.0305}$ & $0.11$ \medskip\\
c121 & $5.90^{+0.027}_{-0.026}$ & $3.32^{+0.0294}_{-0.0305}$ & $0.48$ \medskip\\
c122 & $5.90^{+0.026}_{-0.028}$ & $3.31^{+0.0328}_{-0.0294}$ & $0.47$ \medskip\\
\hline
\end{tabular}
\caption{\label{tab:fit} Best-fit values of the two parameters of the analytic formula for the void spin distributions, 
and the signal to noise ratio of the difference in the void spin distribution between each non-Plank and the Planck models.}
\end{table}

To see if and how the shape and behavior of the void spin distribution, $p(j)$, depends on $\sigma_{8}$,  
we repeat the same analysis with the simulations of four different $\Lambda$CDM models (c113,c116,c130, and c133), the $\sigma_{8}$ values of which are 
different from the Planck value, while all of the other cosmological parameters are fixed at the same Planck values. 
In the top-left panel of figure~\ref{fig:loglog} are shown the resulting void spin distributions with Poisson errors for the four different cases of $\sigma_{8}$ (solid lines) 
and for the case of the Planck $\Lambda$CDM case (dashed line) for comparison. As can be seen, the void spin distributions show appreciable differences among the four 
$\sigma_{8}$ cases. Figure~\ref{fig:pj_sig8} plots the same as figure~\ref{fig:pj} but for the four different cases of $\sigma_{8}$, revealing good agreements between eq.~(\ref{eqn:pj}) 
and the numerical result for each $\sigma_{8}$ case.

Each panel of figure~\ref{fig:cont_sig8} plots the contours of $68\%$,  $95\%$ and $99\%$ confidence regions (solid, dashed and dotted lines, respectively) 
of $\chi^{2}(k,\theta)$ in the two dimensional space spanned by $\{k,\ \theta\}$ for each case of $\sigma_{8}$, and compares them with those
(red filled contours) of the Planck $\Lambda$CDM case (i.e, $\sigma_{8}=0.811$).  
As can be seen, approximately $10\%$ changes of $\sigma_{8}$ from the Planck value induce highly significant differences in the best-fit values of $\{k,\ \theta\}$, 
explicitly demonstrating how sensitively the shape of the void spin distribution depends on $\sigma_{8}$. 

As a measure of difference in $p(j)$ between the Planck and non-Planck $\sigma_{8}$ cases, we measure the distance, 
$D\equiv \left[(k-k^{\prime})^{2}+(\theta-\theta^{\prime})^{2}\right]^{1/2}$, where $\{k,\ \theta\}$ and $\{k^{\prime},\ \theta^{\prime}\}$ denote 
the best-fit parameters of the generalized Gamma distribution for the two cases, respectively: 
The errors in $D$ can be obtained via the error propagation formula~\cite{lee-etal23}: 
\begin{eqnarray}
\label{eqn:sigd}
\sigma^{2}_{D}&=&\left(\frac{\partial D}{\partial k}\right)^{2}\sigma^{2}_{k} + \left(\frac{\partial D}{\partial \theta}\right)^{2}\sigma^{2}_{\theta} 
+ \left(\frac{\partial D}{\partial k^{\prime}}\right)^{2}\sigma^{2}_{k^{\prime}} + \left(\frac{\partial D}{\partial \theta^{\prime}}\right)^{2}\sigma^{2}_{\theta^{\prime}}\, 
+ \nonumber \\
&& 2\left(\frac{\partial D}{\partial k}\right)\left(\frac{\partial D}{\partial \theta}\right)\sigma_{k\theta} + 
2\left(\frac{\partial D}{\partial k^{\prime}}\right)\left(\frac{\partial D}{\partial \theta^{\prime}}\right)\sigma_{k^{\prime}\theta^{\prime}}\, , 
\end{eqnarray}
where $\{\sigma_{k},\  \sigma_{\theta},\sigma_{k\theta}\}$ and $\{\sigma_{k^{\prime}},\  \sigma_{\theta^{\prime}}, \sigma_{k^{\prime}\theta^{\prime}}\}$ 
represent the marginalized errors of the two parameters and the cross-correlations between them for the Planck and non-Planck models, respectively, 
for the computation of which the maximum likelihood function, $p(-\chi^{2}/2)$, is exclusively used.   
In the fourth column of table~\ref{tab:fit} are listed the signal-to-noise ratios, $D/\sigma_{D}$, for the four $\sigma_{8}$-models, revealing that 
approximately $10\%$ change of $\sigma_{8}$ value induces $4\sigma_{D}$ signal of difference in the void spin distribution. 

We suspect that the effect of $\sigma_{8}$ on the void halo mass distributions should play at least a partial role in generating these differences in $p(j)$ among the 
four $\sigma_{8}$ cases shown in figures~\ref{fig:pj_sig8}-\ref{fig:cont_sig8} as well as in table~\ref{tab:fit}.  What is affected most strongly by the variation of 
$\sigma_{8}$ value is the overall amplitude of void halo mass function. Meanwhile we have determined in the current analysis the {\it probability density} function whose amplitude 
is fixed to meet the normalization condition of $\int_{0}^{\infty}\,p(j)\,dj=1$. Nevertheless, the variation of $\sigma_{8}$ also can change the shape of void halo mass distribution 
in the high-mass section where the void halos are very rare. 
In figure~\ref{fig:mass_dis} is shown the {\it normalized} probability density distribution of void halo masses, $m_{h}$ in unit of $h^{-1}\,M_{\odot}$, for the five cases of $\sigma_{8}$. 
The significant differences in $p(\log m_{h})$ among the five $\sigma_{8}$ models are found in the logarithmic mass range of $\log m_{h}\ge 12.5$ where 
$p(\log m_{h})$ drops below $0.1$. 

\subsection{Dependences of the void spin distribution on $\Omega_{\rm cdm}h^{2}$, $M_{\nu}$ and $w$}\label{sec:other} 

Carrying out the same analysis but with the AbacusSummit simulations of four different $\Omega_{\rm cdm}h^{2}$ cases (c102, c103, c131 and c134), 
of two different  ${\nu}$CDM cases  (c009 and c019) and of four $w$CDM cosmologies  (c108,c109,c121 and c122), 
we determine $p(\log m_{h})$ and $p(j)$ for each case, the results of which are shown in the top-right, bottom-left and bottom right panels of figures~\ref{fig:loglog}-\ref{fig:mass_dis}. 
It is worth emphasizing here that when one of the three parameters, $\{\Omega_{\rm cdm}h^{2}, M_{\nu}, w\}$, varies, the other parameters as well as $\sigma_{8}$ 
are fixed at the same values.
As can be seen, both of the void spin and halo mass distributions exhibit much weaker dependences on $\Omega_{\rm cdm}h^{2}$, $M_{\nu}$ and $w$ 
than on $\sigma_{8}$. 

We fit the numerical obtained $p(j)$ to the analytic formula given in eq.~(\ref{eqn:pj}) and quantitatively measure the sensitivity of $p(j)$ to $\Omega_{\rm cdm}h^{2}$, $M_{\nu}$ 
and $w$ by eq.~(\ref{eqn:sigd}), the results of which are shown in figures~\ref{fig:pj_om}-~\ref{fig:cont_w} and in table~\ref{tab:fit}.  
As can be seen, for all of the ten cases, the generalized Gamma distribution indeed describes well $p(j)$ except for in the range of $j\le 1$. 
No statistically significant difference in the best-fit value of $\{k,\theta\}$ is produced even by $~10\%$ change of each parameter.
Varying the values of halo number and mass cuts, we also repeat the analyses and confirm that these results shown in figures~\ref{fig:loglog}-\ref{fig:cont_w}, i.e., 
sensitive dependence of the void spin distribution only on $\sigma_{8}$ is quite robust against the variation of number and mass cuts of void halos.
\begin{figure}[tbp]
\centering 
\includegraphics[width=1.1\textwidth=0 380 0 200]{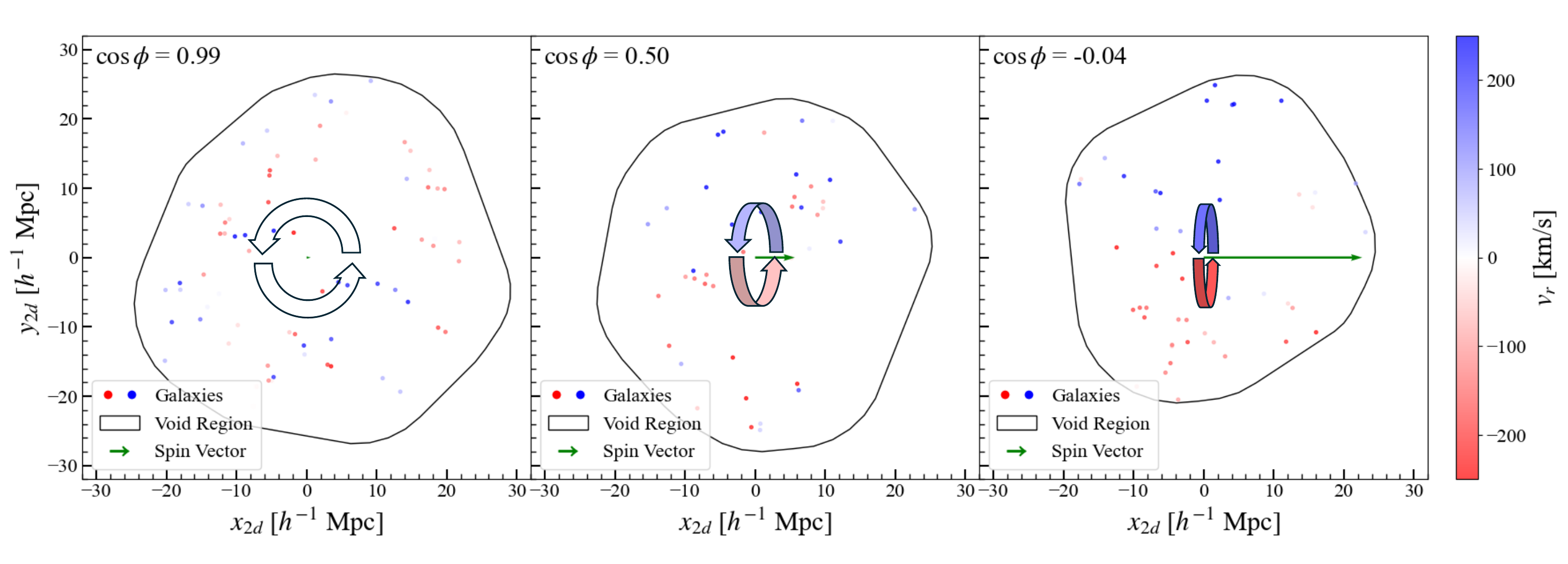}
\caption{\label{fig:proj_sing} Residual values of the radial components of peculiar velocities of void halos in two dimensional 
plane orthogonal to the line of sight direction, $\hat{\bf z}$. Blue dots correspond to the void halos moving toward us 
while the red dots correspond to the void halos moving away from us, due to their spinning motions.}
\end{figure}
\begin{figure}[tbp]
\centering 
\includegraphics[width=1.1\textwidth=0 380 0 200]{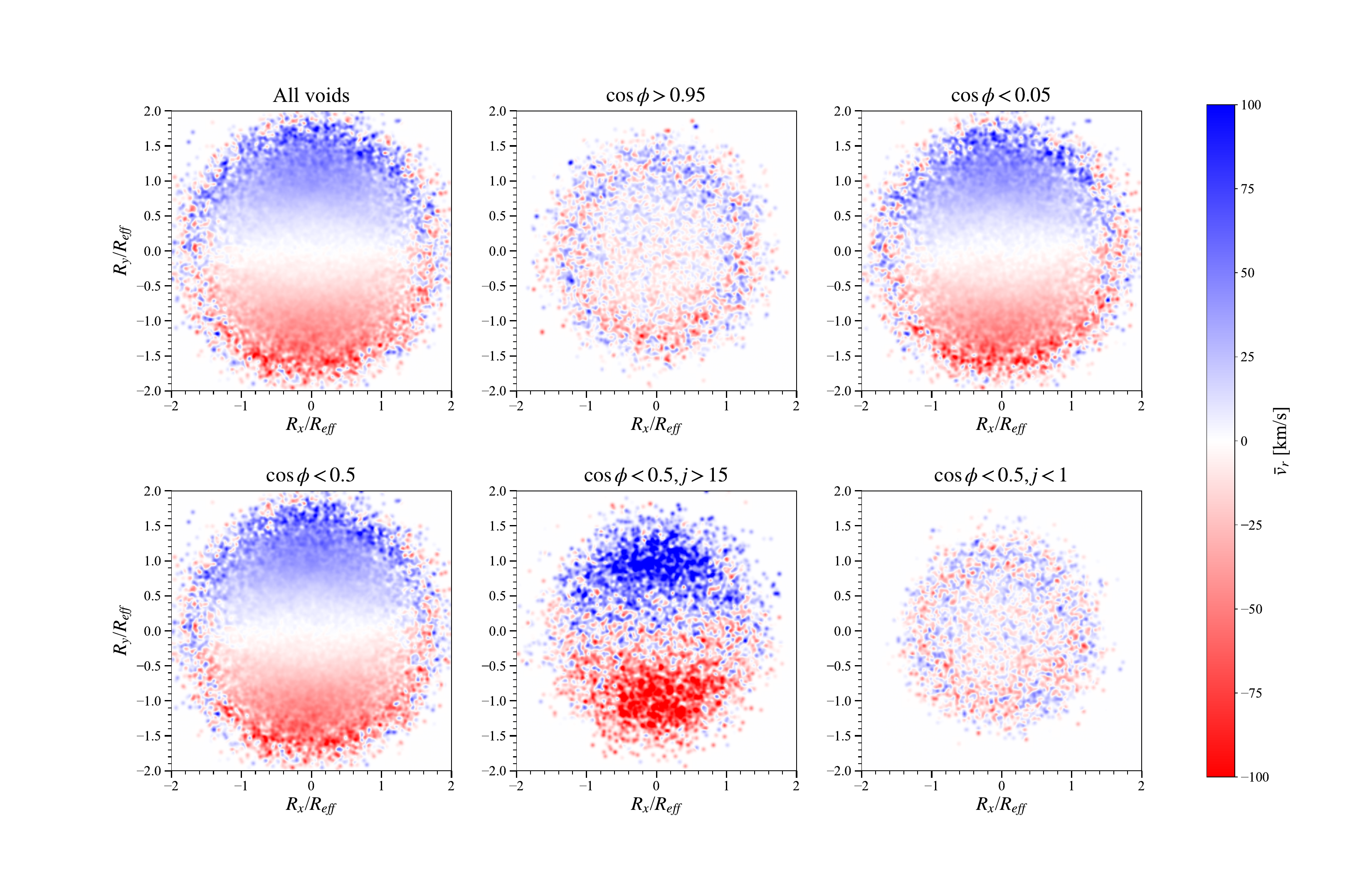}
\caption{\label{fig:red_blue} Residual redshifts and blueshifts of void halos stacked over all voids (top-left panel), 
over the voids whose spin vectors are almost perfectly aligned with the line of sight directions (top-middle panel), over the voids whose spin vectors are almost 
perfectly perpendicular to the line of sight directions (top-right panel), over the voids whose spin vectors are on average aligned with the line of sight directions 
(bottom-left panel), over the voids whose spin vectors have alignment tendency and higher magnitude than unity (bottom-middle panel), over the voids whose 
spin vectors have alignment tendency and lower magnitudes than the unity (bottom-right panel).}
\end{figure}

\section{Observational Feasibility}\label{sec:obs}

Now that the potential of void spin distributions as a sensitive probe of $\sigma_{8}$ is verified,  we would like to address a critical issue of how feasible 
it will be in practice to estimate the void spins from real data.  The major impediment to observational determination of void spin distributions should stem
from large uncertainties involved in the measurements of peculiar velocities of void galaxies. To hurdle this, it would be desirable to develop 
a methodology with which the void spins can be observationally estimated even when information only on the redshifts of void galaxies are available. 

Recollecting the methodology proposed by P.~Wang {\it et al.}~\cite{wan-etal21} to statistically determine the filament spins by measuring only the redshifts 
of filament galaxies, we suggest that the same methodology should also be applicable to the voids, allowing us to determine the void spin distributions only from 
information on their redshifts.  To back up this claim, we evaluate the expected redshifts of the member halos of spinning voids relative to the redshifts of void centers 
by taking the following steps~\cite{wan-etal21}, under the assumption that the $\hat{\bf z}$-axis is parallel to the line of sight direction.
\begin{itemize}
\item
For each void, project the positions of its member halos as well as its spin vector onto the $\hat{\bf x}$-$\hat{\bf y}$ plane normal to the $\hat{\bf z}$ axis,
\item
Compute the radial components of the velocity vectors of void halos as ${\bf v}_{r}\equiv \hat{\bf z}\cdot{\bf v}$, 
and take the average, $\langle v_{r}\rangle$, over all member halos of a given void.
\item
Provided that the spin direction of a given void is not perfectly aligned with $\hat{\bf z}$, the projected spin vector, ${\bf j}_{2d}\equiv {\bf j}\cdot\hat{\bf z}$, 
would divide the projected void region into two sectors: one that contains those halos which exhibit $\Delta v_{r}\equiv v_{r}-\langle v_{r}\rangle>0$ (blueshifts), 
while the other that contains those halos with $\Delta v_{r}\equiv v_{r}-\langle v_{r}\rangle<0$ (redshifts). 
\item
Reorient the coordinate axes of the projected space to have the direction of ${\bf j}_{2d}$ align with its $x$-axis. 
The void halos belonging to the region with $y$-axis coordinate of $y>0$ ($y<0$) will exhibit blueshifts (redshifts). 
\end{itemize}

Figure~\ref{fig:proj_sing} illustrates three projected voids each of which is divided into two sectors by the projected spin vectors (green arrows) aligned with the $x$-axis, 
for three different cases of $\cos\phi$ defined as the dot-product between the void spin vectors and line of sight directions. 
As can be seen,  for the case of $\cos\phi\approx 1$ (i.e., the case of a perfect alignment between ${\bf j}$ and $\hat{\bf z}$, 
the radial velocities of void halos appear to be randomly distributed, showing no mean difference between the two sectors ($y>0$ and $y<0$).  
Whereas, for the case of  of $\cos\phi < 1$, the radial velocities of void halos exhibit two distinct behaviors, {\it blueshifts and redshifts}, between the two sectors, 
due to the void spinning motion (curved arrows). The smaller the value of $\cos\phi$ is, the larger the difference between the blueshift and redshift behaviors of the void halos 
belonging to the two sectors. 

Figure~\ref{fig:red_blue} depicts the expected residual redshifts and blueshifts of the member halos stacked over various voids in the plane normal to $\hat{\bf z}$. 
The $x$ and $y$ components of the radial separation vectors of the stacked void halos, $(R_{x},R_{y})$,  from the void centers are rescaled by the effective void radii, $R_{\rm eff}$. 
As can been, even in the cases that all voids are stacked (top-left panel), we detect a clear signal of the residual redshifts and blueshifts of the member halos of spinning voids. 
When the stacking is made over the voids whose spin axes have lower degree of alignments with the line-of-sight directions ($\cos\phi<0.5$) and very high magnitudes 
($j\ge 15$), we witness the highest signal of the differences between the redshifts and blueshifts of the void halos located in the dichotomized sections of the voids 
(bottom middle panel).  No signal of the residual redshift and blueshift is found for the opposite case of ($j< 1$) (bottom right panel). 
This feasibility test implies that the probability density distribution of projected void spins, $j_{2d}$, can in principle be obtained by measuring the maximum difference between the blueshifts and redshifts of the void galaxies without having any information on their peculiar velocities.

In reality, of course, there is no available information on the directions of ${\bf j}_{2d}$, which plays the vital role of dichotomizing the void regions into two distinct 
sectors where the void galaxies yield the maximum blueshift and redshift differences. 
As done in the work of P.~Wang {\it et al.}~\cite{wan-etal21}, it should be possible to find the projected void spin direction through random optimization process: 
Iteratively creating a two dimensional direction, computing the blueshift-redshift difference between the two regions dichotomized by the randomly created directions, 
until the best direction which yields the maximum blueshift and redshift difference is found. This optimization process is, however, beyond the scope of this paper. 


 \section{Summary and Discussion}\label{sec:con}
 
With the help of the AbacusSummit simulations which made it achievable to investigate the single parameter dependence of any late-time probe, 
we have numerically discovered that the void spin distribution has a potential to diminish the degeneracy of $\sigma_{8}$ with $\Omega_{\rm cdm}h^{2}$, 
$M_{\nu}$ and $w$.  From each of $15$ different AbacusSummit simulations whose initial conditions are different among one another only in one of the aforementioned 
four cosmological parameters, the voids have been identified via the Void-Finder algorithm~\cite{HV02} which has an advantage of being directly applicable to the 
spatial distribution of halos without requiring any assumption on the matter field.  The magnitude of the rescaled specific angular momentum of each identified void has been 
determined as a void spin by treating its member halos as equal mass point-like objects. 
The cosmology dependence of the void spin distribution has been expected on the grounds that the member halos of a void must 
possess not only radial but also tangential motions relative to its center~\cite{sha-etal06} due to the anisotropic tidal field of surrounding cosmic web~\cite{cosmic_web}.
For the void spin distribution of each model, we have included only those voids containing $15$ or more well-resolved halos with masses 
$\ge 10^{11.5}\,h^{-1}M_{\odot}$.

We have also for the first time shown that the void spin distributions are excellently described by the generalized Gamma distribution 
characterized by two adjustable parameters, $\{k, \theta\}$ for all of the $15$ cosmologies considered in the current work. 
The best-fit values of $\{k,\theta\}$ have been determined via the $\chi^{2}$-statistics and shown to vary significantly only with $\sigma_{8}$. 
Quantifying the differences in the void spin distributions between the Planck $\Lambda$CDM and other cosmological models 
by the distances between the best-fit parameter vectors, we have found that $\sim 10\%$ deviation of $\sigma_{8}$  from the Planck value~\cite{planck18} induces a  difference 
in the best-fit values of $\{k,\theta\}$ as significant as $\sim 4\sigma$. Meanwhile, no such sensitive dependence of of $\{k,\theta\}$ on $\Omega_{\rm cdm}h^{2}$, $M_{\nu}$ and 
$w$ have been witnessed.  

It has also been examined how feasible it would be to measure the void spins from real data when only the redshifts of void galaxies are available.
Applying to the voids the scheme devised in the heuristic work of P.~Wang {\it et al.}~\cite{wan-etal21} to detect the spinning motions of cosmic filaments, 
we have shown that the void spins could be well estimated in the plane of sky by measuring only the redshift differences of void galaxies. 
This scheme has been found to be much more feasible if we include only those voids whose spins have directions misaligned with the line of sight and magnitudes  
high enough to generate appreciable redshift differences between the member galaxies belonging to the approaching and receding sectors 
dichotomized by the projected void spin directions. 

Now that the auspicious outcome of the feasibility test has enhanced a prospect of void spin distribution as a complimentary probe of $\sigma_{8}$, 
we would like to discuss several caveats and limitations of this new diagnostics, suggesting that some future tasks be carried out to overcome the limitations.
First, in the current analysis have been included only those giant voids having $15$ or more member halos, identified via the classical Void-Finder algorithm~\cite{HV02}. 
It may not necessarily be a downside of the current analysis given that the void spins can be more accurately measured for the case of giant voids having larger 
number of void halos and that the giant voids are more susceptible to the anisotropic tidal effects of large-scale matter distribution, reflecting better the initial conditions~\cite{LP06}. 
Nevertheless, it would be definitely desirable to explore if it would help putting more precise constraints on $\sigma_{8}$ to include smaller voids with fewer tracers by testing 
the robustness of void spin distributions against the variation of void-finding algorithms.

Second, it has yet to be understood how closely the $\sigma_{8}$-effect on the void halo mass distribution is related to the sensitivity of the void spin distribution to $\sigma_{8}$ 
found in the current analysis. We have witnessed that both of the void spin and halo mass distributions are dependent on $\sigma_{8}$ while both are almost independent on the 
other three parameters (figures~\ref{fig:loglog}-\ref{fig:mass_dis}). 
Although the void halos have been treated as equal-mass point-like components in the current analysis, the determination of void spin distributions still requires knowledge 
on the total halo masses of individual voids for the measurements of the circular velocity $V_{v}$ that appears in the definition of void spins, eq.~(\ref{eqn:def_vspin}).  
It will be definitely necessary with the help of information theory~\cite{sha48} to properly assess how much independent and complementary information the void spin distributions contain 
about $\sigma_{8}$ compared with the other void statistics based on the first and second properties of void halos (i.e., positions and velocities) like the void shapes, sizes and velocity profiles 
as well as the void halo abundances~\cite{LP09,bis-etal10,bos-etal12,LW12,mas-etal15,NT17,ver-etal19,ham-etal20,rez20,dav-etal21,con-etal22,ver-etal23,ebr24}.   

Third, successful as the generalized Gamma distribution has been found in describing the void spin distributions for all of the cosmological models considered, 
it will be highly desirable to derive a truly physical model for the void spin distributions from the first principles expressed in terms of initial conditions. 
To develop such a physical, it will be necessary to redefine the void spins in terms of DM particles rather than the void halos, which will in turn require to take into proper 
account the matter-to-halo bias in void environments.

Fourth, it will be necessary to trace the evolution of the void spin distribution by utilizing light cone simulations and to investigate if it still retains the sensitive 
$\sigma_{8}$-dependence even at higher redshifts, where a larger number of voids are available. Tracing the redshift evolution of void spin distribution will also 
allow us to investigate more completely how the time varying DE equation of state affects the void spin distribution. Moreover, it will provide information on the other key 
cosmological parameters since the $\sigma_{8}$ values measured at higher redshifts depend on the other parameters.
Fifth, a more thorough investigation of the $M_{\nu}$ dependence of the void spin distribution is required to conclude that this diagnostic is indeed capable of 
breaking the $\sigma_{8}$-$M_{\nu}$ degeneracy by considering the full nonlinear effects of massive neutrinos as gravitating particles.
Our future work will be in the direction of making these improvements.

\acknowledgments
JL acknowledges the supports by Basic Science Research Program through the NRF of Korea funded by the Ministry of Education (No.2019R1A2C1083855 and 
RS-2025-00512997).

\end{document}